\documentclass[fleqn,usenatbib]{mnras}

\usepackage{newtxtext}

\usepackage[T1]{fontenc}

\DeclareRobustCommand{\VAN}[3]{#2}
\let\VANthebibliography\thebibliography
\def\thebibliography{\DeclareRobustCommand{\VAN}[3]{##3}\VANthebibliography}

\usepackage{graphicx}	
\usepackage{amsmath}	
\usepackage{amssymb}	
\usepackage{float}
\usepackage{array} 
\usepackage{xspace}    

\usepackage{diagbox}  
\usepackage{arydshln} 
\usepackage{multirow} 

\newcommand{\moy}[1]{\left\langle#1\right\rangle}  
\newcommand{\Var}[2][]{\mathrm{Var}_{#1}[#2]}
\newcommand{\RE}[1]{\operatorname{\Re}\left\{#1\right\}}

\newcommand{\Cov}[1]{\mathrm{Cov}[#1]}
\renewcommand{\micron}{\textmu{}m\xspace}   
\newcommand{\ds}{\,\mathrm{d}\sigma}
\newcommand{\NOTE}[1]{\textit{\footnotesize [NB: #1]}}
\renewcommand{\NOTE}[1]{}  
\newcommand{\NotE}[1]{\textcolor{red}{\textit{\footnotesize [MODIF: #1]}}}
\renewcommand{\NotE}[1]{}  

\title[Multiband differential dispersion correction]{Compensation of differential dispersion: application to multiband stellar interferometry.}

\author[C. Pannetier et al.]{
Cyril Pannetier$^{1,2}$\thanks{E-mail: cyril.pannetier@oca.eu (CP)}, Denis Mourard$^{1}$, Frédéric Cassaing$^{2,3}$, Stéphane Lagarde$^{1}$, Jean-Baptiste Le Bouquin$^{4}$, \and John Monnier$^{5}$, Judit Sturmann$^{6}$, and Theo Ten Brummelaar$^{6}$.
\\
$^{1}$Université Côte d'Azur, Observatoire de la Côte d'Azur, CNRS, Laboratoire Lagrange, Bd de l'Observatoire, CS 34229, 06304 Nice cedex 4, France\\
$^{2}$DOTA, ONERA, Université Paris Saclay, F-92322 Châtillon, France\\
$^{3}$Now at DTIS, ONERA, Université Paris Saclay, F-91123 Palaiseau, France\\
$^{4}$Univ. Grenoble Alpes, CNRS, IPAG, F-38000 Grenoble, France\\
$^{5}$University of Michigan, Ann Arbor, MI 48109 USA\\
$^{6}$CHARA Array, Mount Wilson Observatory, 91023 Mount Wilson CA, USA
}

\date{Accepted XXX. Received YYY; in original form ZZZ}

\pubyear{2021}

\begin{document}
\label{firstpage}
\pagerange{\pageref{firstpage}--\pageref{lastpage}}
\maketitle

\begin{abstract}
With the aim of pushing the limiting magnitude of interferometric instruments, the need for wide-band detection channels and for a coordinated operation of different instruments has considerably grown in the field of long-baseline interferometry. For this reason, the Center for High Angular Resolution Astronomy (CHARA), an array of six telescopes, requires a new configuration of longitudinal dispersion compensators to keep the fringe contrast above 95\% simultaneously in all spectral bands, while preserving the transmission above 85~\%. In this paper, we propose a new method for defining the longitudinal dispersion compensators (LDC) suited for multi-band observations. A literal approximation of the contrast loss resulting from the dispersion residues enables us to define a general criterion for fringe contrast maximisation on several bands simultaneously. The optimization of this criterion leads to a simple solution with only two LDC stages per arm and existing differential delay lines, to the glass choice and a simple linear formula for thickness control of all these media. A refined criterion can also take into account glass transmission. After presenting this criterion, we give the optimal solution (medium, configuration) and its expected performance for the planned observing modes on CHARA.
\end{abstract}




\section{Introduction}
Ground-based long-baseline optical interferometers combine the light from an array of telescopes distant by tens to hundreds of meters. The combination of the electromagnetic fields creates an interferometric pattern whose characteristics depend on the interferometer configuration and the source properties. It gives access to the angular resolution of a single telescope of diameter equal to the maximum distance between the telescopes in the array. Because they are operated on broadband sources, performance of these  interferometers intended to measure the spatial coherence is critically affected by temporal coherence. The current generation of optical long-baseline interferometers, schematically represented in Fig. \ref{fig:geometric_delay}, uses main delay lines (MDL) to equalize the optical paths, which achromatic difference is mostly introduced in vacuum outside the atmosphere. However, if this equalization is done in air, the dispersion law of this medium introduces a dispersion at the interferometric focus. The equalization of the optical path becomes chromatic and not only geometric: the coherence envelop multiplying the interferometric fringes is chromatically shifted and the fringes are blurred within a spectral channel as it gets wider.

This problem has been identified and solved from the very first stellar interferometer with the introduction of LDC, usually made of pieces of glass of variable thickness to compensate the air thickness \citep{labeyrie-interfVega-aj75}. Another solution, currently being implemented at Magdalena Ridge Optical Interferometer \citep{creech-eakman_magdalena_2018}, is to use evacuated delay lines.

The recent years have seen important developments of the interferometric instrumentation \citep{gravity2017, MatisseMsgr, MIRCX2020, monnier_mystic_2018, mourard_spica_2018}. For measuring visibilities on fainter objects, the need for large spectral channels instruments \citep{gravity2017, mourard_spica_2018} and for a coordinated operation of different instruments \citep{Matter2010, mourard_spica_2017, MatisseMsgr} has considerably grown. Indeed, this coordination not only enables to track fringes in one band while integrating in all other bands but also increases the available observing time of all instruments. For example, the Gravity Fringe Tracker~\citep{lacour_gravity_2019} working in the K band is now used to stabilize the fringes of the instrument MATISSE~\citep{MatisseMsgr} in the L and M band, with the introduction of a correction of the chromatic group-delay with the internal delay lines of MATISSE.

CHARA is located at Mount Wilson (California, USA). It has been designed by the University of Georgia and entered into service in 1999. Its six 1-meter telescopes are distant from each others by 33 to 330~meters. From 2007, the VEGA instrument was used in the R-band~\citep{mourard_performance_2012} in medium to high spectral resolution. Another instrument, CLASSIC/CLIMB~\citep{ten_brummelaar_classicclimb_2013}, observes mainly in the K band at low spectral resolution. These two instruments can operate together thanks to the first and still in place LDC in CHARA~\citep{berger_preliminary_2003}.
But a new generation of instruments has been arriving on CHARA since 2018. CHARA/SPICA (Stellar Parameters and Images with a Cophased Array, \cite{mourard_spica_2018}), is an instrument operating mainly in low spectral resolution mode (R=140) over a large band in the visible range from 0.6 to 0.9~\micron. It is assisted by a fringe tracker using a low spectral resolution (R=20) sensor in the H band (1.65~\micron) controlling the fast stages of the main delay lines. Moreover, the coordinated operation of SPICA in the visible, MIRCx~\citep{MIRCX2020} (and the fringe sensor) at R=20 in the H-band, as well as MYSTIC~\citep{monnier_mystic_2018} at R=20 in the K band is considered, leading to a very wide interferometric band from 0.6 to 2.45~\micron.
However, the current implementation of the LDC in CHARA presents two main limitations. First, it implies a transmission loss of 1 magnitude in the K-band. Second, for the planned simultaneous observations in the four bands at low spectral resolutions, this solution can’t reach the level of dispersion correction required by the instruments as we will see in Sect.~\ref{subsec:LDCconfiguration}. Thus, it is clear that a new optimised LDC solution is mandatory.

The longitudinal dispersion compensation has been very well described by \citet{tango_dispersion_1990} with the introduction of as many LDCs as necessary for an optimized correction. However, it turns out that the solutions that have been proposed and implemented~\citep{ten_brummelaar_differential_1995,lawson_dispersion_1996,berger_preliminary_2003} have always simplified the problem because of practical considerations (limited number of LDCs, limited number of spectral bands).
On CHARA, we need to maximise the fringe contrast on up to four separated spectral bands simultaneously. However, considering the available space on the optical tables and the usual transmission in the K band of glasses well adapted to the correction of dispersion, we decided to limit the correction to two stages. This prevents us from optimising each band individually as do the approaches relying on Tango's formalism. In this paper, we propose a rewriting of the criterion for the minimization of the dispersion on as much spectral bands as wanted for a given number of LDCs. It permits us to achieve the best overall performance that the set of LDCs in presence can reach.

We first develop in Sect.~\ref{sec:theory} the new theoretical approach of longitudinal dispersion compensation. Then, in Sect.~\ref{sec:solution}, we present the optimal solution and its expected performance for the planned observing modes on CHARA. A discussion in Sect.~\ref{sec:discussion} on the possible other applications of this study and the future limits of dispersion compensation on long-baseline interferometers is proposed.

\section{Dispersion compensation}
\label{sec:theory}

\subsection{Problematic}
\label{sec:Problematic}

In stellar interferometry, the phase-delay is the phase difference between the electromagnetic fields coming from the two arms and recorded by the detector. The group-delay (GD), proportional to the chromatic gradient of the phase-delay, is what governs the contrast of the interferometric fringes. These two quantities will be expressed mathematically in the subsection \ref{sec:Tango}. When the GD is null, the phase-delay is locally achromatic and all the fringes at different wavelengths in the considered spectral channel superimpose such that the polychromatic fringe contrast is maximum. This is the zero group-delay (ZGD) position. However, since the LDC can't match perfectly the air dispersion, this local achromatism may not be maintained on a large waveband and the group-delay can't be nulled for all wavelengths. Therefore, the residual chromatism within a given spectral band generates a loss of contrast due to the blurring caused by the non-coherent addition of fringes of different wavelengths. Considering the needs for sensitivity (large spectral bands, increasing the limiting magnitude) and for the optimal use of the large interferometric arrays, it is therefore critical to correct this effect.

\begin{figure}
\includegraphics[width=\columnwidth]{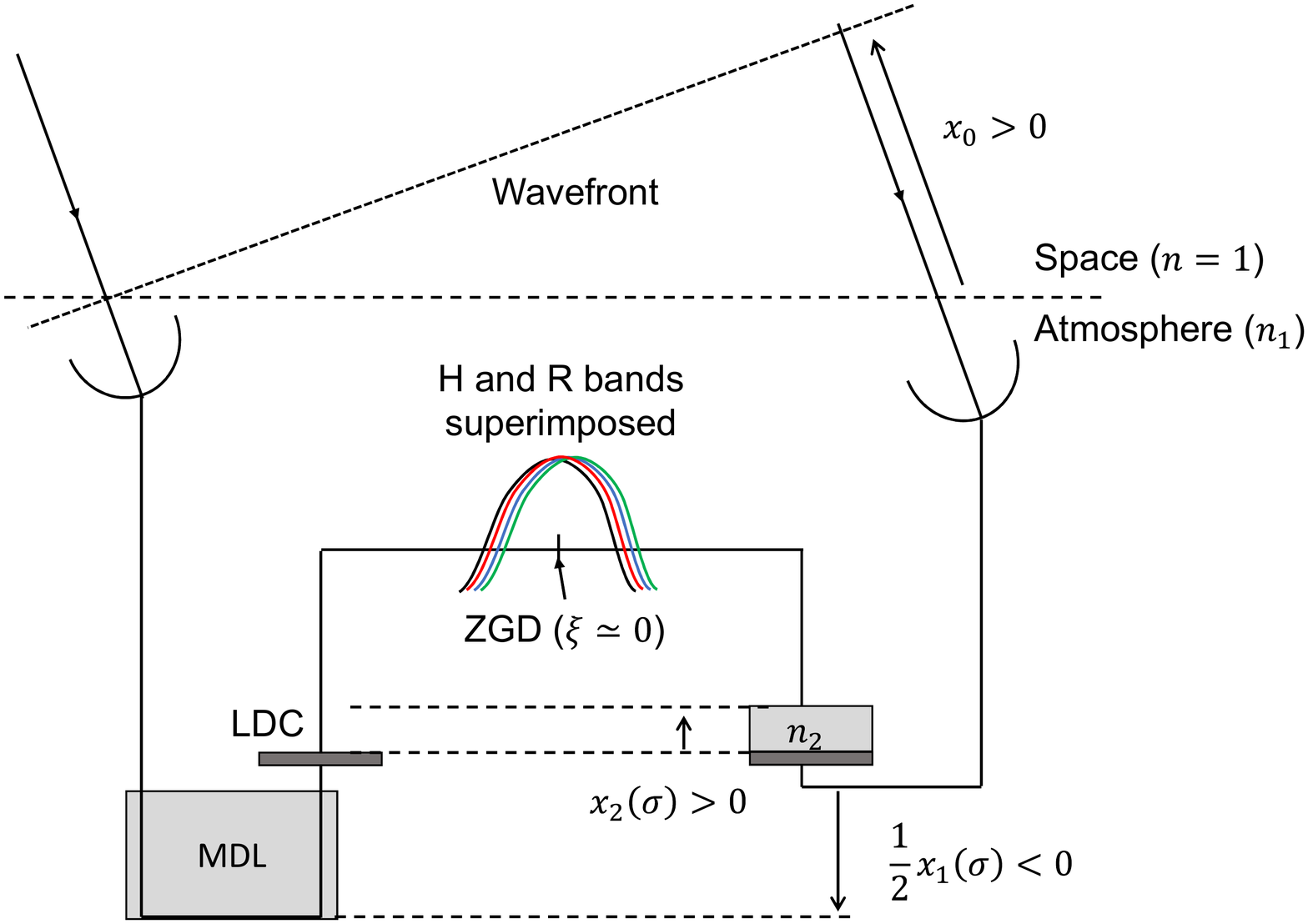}
\caption{Representation of the geometric delay ($x_0$) and the correction of the chromatic delay between the H and R bands on a ground-based interferometer. The chromatism of the optical delay lines correction ($x_1(\sigma)$) is compensated with the chromatism of the LDC ($x_2(\sigma)$) made of a suitable glass with thickness such that the group-delay of all spectral channels is close to the one of the tracking band, resulting in the superimposition of the coherence envelops.}
\label{fig:geometric_delay}
\end{figure}

It is important to note that no glass presents exactly the same dispersion law as the air. Therefore, each additional medium enables to improve the coherent addition of the fringes inside a given spectral band only. At the same time, the MDL shifts the fringes to the ZGD by nulling the residual group-delay introduced by the LDCs.
After compensation with the LDCs, the interferometer modulates the fringes to observe them.

Our goal is to maximize the fringe contrast on the detector, while keeping the highest possible transmission. The purpose of the study presented in this paper is to determine the nature and thickness of the glasses mounted on the LDCs.
We neglect the atmospheric phase disturbance and we present the formalism in a 2-beam configuration, the extension to N-beam being straightforward. Finally, we assume unit visibility fringes for simplicity.


\subsection{Classical solution}
\label{sec:Tango}

We consider two beams crossing $N$ media of dispersion law $n_i(\sigma)$. We define the optical path difference (OPD) as:
\begin{equation}
    X(\sigma)=\sum_{i}n_i(\sigma)x_i
    \label{eq:OPD}
\end{equation}
where the wavenumber $\sigma$ is the inverse of the wavelength of the electromagnetic field and $x_i$ is the algebraic differential thickness of the $i$-th medium between the two arms. By convention, an excess thickness is positive when it is in the same arm than the geometrical delay noted $x_0$. The differential thickness of the MDL is noted $x_1$. A simplified illustration is presented in Fig.~\ref{fig:geometric_delay}.

The phase-delay is defined by:
\begin{equation}
    \Phi(\sigma)=2\pi\sigma X(\sigma)
    \label{eq:PhaseDelay}
\end{equation}

The group-delay is defined as the gradient of the phase-delay, expressed as a vacuum length for convenience:
\begin{equation}
    \xi(\sigma)=\frac{1}{2\pi}\dfrac{d\Phi}{d\sigma}(\sigma)
    \label{eq:GroupDelay_Gradient}
\end{equation}

which leads to the simple expression:
\begin{equation}
    \xi(\sigma)=\displaystyle\sum_i n_{gi}(\sigma)x_i
    \label{eq:GroupDelay_Index}
\end{equation}
where $n_{gi}$ is the group index of the $i$-th medium, defined as:
\begin{equation}
    n_{gi}(\sigma) = n_i(\sigma)+\sigma\frac{dn_i}{d\sigma}(\sigma)
    \label{eq:GroupIndex}
\end{equation}

Under the condition of a low phase variation $\Phi$ over a small spectral distance $s$ from the mean wavenumber $\bar{\sigma}$, the Taylor expansion of the phsae-delay is:
\begin{equation}
    \Phi(\bar{\sigma}+s) = \Phi(\bar{\sigma}) + 2\pi[(x_0 + \mathbf{b}_1.\mathbf{x})s + \mathbf{b}_2.\mathbf{x}s^2+\mathbf{b}_3.\mathbf{x}s^3+O(s^4)],
    \label{eq:TaylorExpansion}
\end{equation}

with $O$ being the asymptotic notation and where the vectors $(\mathbf{b}_j)_{j\in\mathbb{N}}$ are such that the $i^{th}$ component of the vector $\mathbf{b}_j$ is the $j$\textsuperscript{th} coefficient of the Taylor expansion of the quantity $\sigma n_i(\sigma)$. We identify three different terms:
\begin{itemize}
    \item the phase-delay $\Phi(\bar{\sigma})$ at the wavenumber $\bar{\sigma}$,
    \item the group-delay at the wavenumber $\bar{\sigma}$ responsible for the fringe's position shift:
    \begin{equation}
        \check{\xi}=x_0 + \mathbf{b}_1.\mathbf{x}
        \label{eq:GroupDelay_Tango}
    \end{equation}
    \item the pure dispersion function:
    \begin{equation}\label{eq-pure-disp}
        \check{\Psi}(s)=2\pi(\mathbf{b}_2.\mathbf{x}s^2+\mathbf{b}_3.\mathbf{x}s^3)+O(s^4)
    \end{equation}
    responsible for the residual fringe contrast within a spectral channel.
\end{itemize}

\cite{tango_dispersion_1990} demonstrates that the fringe contrast at the ZGD, resulting from the pure dispersion function on a spectral channel $\Sigma$, can be approximated by:
\begin{equation}
    C_{\Psi,\Sigma}=C_{\Phi,\Sigma}(\xi=0) \simeq 1-\Var{\check{\Psi}}.
    \label{eq:TangoContrast}
\end{equation}

From the equation~(\ref{eq:TangoContrast}), Tango concludes that maximizing the fringe contrast at the ZGD on a given spectral channel means minimizing the variance of the residual dispersion.

After development of the standard deviation of the dispersion function $\Psi$, and without forgetting the null group-delay condition, it turns out that to maximize the fringe contrast we need to verify by order of priority the following equations:
\begin{align}
     \mathbf{b}_1.\mathbf{x} &=-x_0 \nonumber\\
     \mathbf{b}_2.\mathbf{x} &= 0 \nonumber\\
     \mathbf{b}_3.\mathbf{x} &= 0\label{eq:conditions}\\
     ...&\nonumber\\
     \mathbf{b}_N.\mathbf{x} &= 0\nonumber
\end{align}

This set of conditions proposed by \cite{tango_dispersion_1990} is very convenient both for its simplicity of use and its clarity. Albeit suitable for maximising the fringe contrast on a given spectral band, it is no longer adapted to a wide band with many sub-channels and different resolutions. Indeed, let's assume that we want to observe with four instruments distributed on four different bands, each one equipped with a differential delay line (DDL). In addition to these DDL, we add LDCs. To optimize the fringe contrast for all instruments together, we thus need to verify for each spectral band as many equations as possible. The common and differential optical delay lines enable us to null the group-delay (first equation) at the center of each band. Then, with two LDCs, we can null the first equation of only two of the four bands. Adding more LDCs may unfortunately generate important transmission loss. So if we want to solve our multi-bands problem, we need another criterion.

\subsection{A method for multi-bands fringe contrast maximization}
\label{sec:MultiBand}

A solution would be to minimize the equations of the system~(\ref{eq:conditions}) rather than null them. In order to get a low dispersion at the center of all bands, we minimize the expression~(\ref{eq-pure-disp}). Weights with positive reals ($W_1,W_2,W_3,W_4$) could also be introduced in this minimisation to weight differently the spectral bands.
\begin{equation}
    L(\mathbf{x}) = W_1(\mathbf{b}_{12}.\mathbf{x})^2+W_2(\mathbf{b}_{22}.\mathbf{x})^2+W_3(\mathbf{b}_{32}.\mathbf{x})^2+W_4(\mathbf{b}_{42}.\mathbf{x})^2,
\end{equation}
where $\mathbf{b}_{\Sigma 2}$ is the second Taylor coefficients vector of the $\Sigma$\textsuperscript{th} spectral band.
Doing that, we can maximise the fringe contrast at the center of each band. However, the use of the Taylor coefficients $\mathbf{b}_i$ makes it necessary to choose one precise wavenumber per spectral band. Thus, we lose information about the rest of the dispersion law and we are not guaranteed to find the best overall contrast. This difference gets larger as the individual spectral bands get wider.

For this reason, we decided to use a maximisation method that takes all the spectral information into account. In particular, we do not distinguish the group-delay from the dispersion function anymore. This new criterion focuses on the overall minimization of the dispersion on the whole bands. It means that we do not try to null all the high-order derivatives at a given somehow arbitrary point like suggests the system~(\ref{eq:conditions}) but rather directly maximize the fringe contrast over the whole band of interest.

We show in appendix~\ref{sec:ContrastLoss} that the fringe contrast of a polychromatic interferogram resulting from the phase-delay $\Phi(\sigma)$ over the spectral channel $\Sigma$ is:
\begin{equation}
    C_{\Phi,\Sigma} \simeq \exp(-\Var[\Sigma]{\Phi}/2)
    \label{eq:FringeContrast}
\end{equation}

The new expression of the fringe contrast in equation~(\ref{eq:FringeContrast}) differs from the equation~(17) of \cite{tango_dispersion_1990} (eq. \ref{eq:TangoContrast} in this paper) by two facts. First, $\Phi$ contains the group-delay counterpart, which enables us to make the overall optimisation that we are looking for. Second, the exponential approximation remains true at a higher order than the 2\textsuperscript{nd} order Taylor expansion in Tango's equation~(17).

For a given phase-delay $\Phi(\sigma)$, the equation~(\ref{eq:FringeContrast}) enables us deriving the associated fringe contrasts $C_{\Phi,\Sigma}$ in each considered spectral channel $\Sigma$.
The goal is to optimize the fringe contrast on as many spectral channels $\Sigma$ as possible. Thus, we need to minimise $L(\mathbf{x})$ defined as: 
\begin{equation} \label{eq:criterelog}
    L(\mathbf{x})=-\sum_{\Sigma}W_\Sigma\log C_\Sigma
\end{equation}
where $W_\Sigma$ are weights making possible to favour an instrument before another. By default, they are all put to 1.

The logarithmic expression, classical for multiplicative losses, is introduced for giving a higher weight to high losses.
The criteria $L(\mathbf{x})$ is a quadratic function of the vector $\mathbf{x}$. This means its minimisation is linear and, as detailed in appendix~\ref{sec:MaximizationEquation}, leads to the linear matrix equation:
\begin{equation}
\mathbf{M} \cdot \mathbf{x}'=\mathbf{d}
\label{eq:minimization_equation}
\end{equation}
where we introduced:
\begin{itemize}
    \item the vector $\mathbf{x}' = (\delta x_1, x_2, x_3, ...,x_N)$ where $x_i$ is the excess thickness of the $i$-th media as already defined before and $\delta x_1$ is the additional thickness of the MDL that adds up to the first order excess thickness 
    \begin{equation}
        x'_1=-\dfrac{x_0}{n_{g,1}(\sigma_0)}
    \end{equation} that corrects the group-delay for an arbitrary wavenumber $\sigma_0$. So $x_1 = x'_1+\delta x_1$.
    \item the vector $\mathbf{d}=(d_i)_{i\in [1,N]}$, made of the covariances between the residual phase dispersion after the first-order correction by the MDL and the dispersion laws of the additional media, defined as
    \begin{equation}
        \begin{aligned}
        d_i=&-x_0 \displaystyle\sum_{\Sigma}W_\Sigma\int_{\Sigma} (\tilde{n}_{\epsilon}(\sigma)-\moy{\tilde{n}_{\epsilon}})(\tilde{n_i}(\sigma)-\moy{\tilde{n_i}})\ds\\
        d_i=&-x_0\displaystyle\sum_{\Sigma}W_\Sigma\Delta\sigma_{\Sigma} \Cov{\tilde{n}_{\epsilon},\tilde{n}_i}_{\Sigma}
        \end{aligned}
    \end{equation}
    \item and the matrix $\mathbf{M}=(m_{ij})_{(i,j)\in[1,N]^2}$, made of the covariances between the dispersion laws of all additional medium, defined as 
    \begin{equation}\label{eq-mij}
        \begin{aligned}
        m_{ij}=&\displaystyle\sum_{\Sigma}W_\Sigma\int_{\Sigma}(\tilde{n_i}(\sigma)-\moy{\tilde{n_i}})(\tilde{n_j}(\sigma)-\moy{\tilde{n_j}})\ds\\
        m_{ij}=&\displaystyle\sum_{\Sigma}W_\Sigma\Delta\sigma_{\Sigma} \Cov{\tilde{n}_i,\tilde{n}_j}_{\Sigma}
        \end{aligned}
    \end{equation}
\end{itemize}

The two last variables also use notations that we need to define:
\begin{itemize}
    \item the quantity $\tilde{n}_i(\sigma)=\Pi_i(\sigma)\sigma n_i(\sigma)$ that concerns the $i$\textsuperscript{th} glass where $\Pi_i(\sigma)$ is a "flag" function equal to 1 on the spectral range that sees the medium and 0 elsewhere. This enables to model the case of the DDL or LDC placed in a specific spectral band.
    \item $n_{\epsilon}(\sigma) = 1 - n_1(\sigma)/n_{g,1}(\sigma_0)$ is the "extra" index of air with respect to vacuum that remains after the correction of the ZGD for the arbitrary wavenumber $\sigma_0$. $\tilde{n}_{\epsilon}$ follows the same definition than $\tilde{n}_i$.
    \item $\moy{.}$ is the chromatic average.
\end{itemize}

To stay as general as possible, we could have kept the MDL length $x_1$ rather than $\delta x_1$ in the vector $\mathbf{x}'$. However, in practice, this approach is sensitive to numerical noise due to the high disproportion between the components of the resulting vector $\mathbf{x}=(x_1,x_2,...,x_N)$. Indeed, in this general approach, the first component $x_1$ of $\mathbf{x}$ is meter-scaled whereas its N-1 last components, corresponding to the medium compensating the residual dispersion, are only millimeter or micrometer scaled. This same disproportion is present in $\mathbf{M}$ and $\mathbf{d}$. The matrix $\mathbf{M}$ is ill-conditioned, leading to numerical errors at its inversion. Introducing this intermediate dispersion correction with the MDL length $x'_1$ enables to put all the dispersion residues at closer scales for every medium. The conditioning of $\mathbf{M}$ doesn't change but the resulting errors are sufficiently low to get robust results.

Yet, as no medium perfectly matches the air dispersion law, the problem is not degenerated. The inverse $\mathbf{M}^{-1}$ of the matrix $\mathbf{M}$ exists and this equation~(\ref{eq:minimization_equation}) admits only one solution that corresponds to the control equation for the $N$ media that form the dispersion control in addition to the first order correction $x'_1$:
\begin{equation}
    \mathbf{x}'_{opt}= \mathbf{M}^{-1}\cdot\mathbf{d}
    \label{eq:MinimizationSolution}
\end{equation}

\subsection{Estimating the final fringe contrast}

The equation~(\ref{eq:FringeContrast}) gives the fringe contrast in a spectral channel $\Sigma$ at the center of the instrument modulation range.
As it is processed from the total phase-delay $\Phi(\sigma)$, it takes into account the fringe contrast due to the group-delay $\xi$.
If the group-delay is higher than the modulation range of the instrument, the equation~(\ref{eq:FringeContrast}) is a good approximation of the fringe contrast measured.
However, if $\xi$ remains into the modulation range, the measurement still benefits from the highest fringe contrast of the coherence envelop. In this case, assuming that the second and higher orders of the dispersion law vary very slowly inside the modulation range, the final fringe contrast is given by:
\begin{equation}
C_{\Psi,\Sigma} \simeq \exp(-\Var[\Sigma]{\Psi_\Sigma}/2)
\label{eq:MaximalFringeContrast}
\end{equation}
where
\begin{equation}
\Psi_\Sigma(\sigma) = \Phi(\sigma) - (\moy{\Phi}_\Sigma + \sigma\Phi'(\sigma))
\label{eq:DispersionFunction}
\end{equation}
is the pure dispersion function analog to $\check{\Psi}$ defined in the equation~\ref{eq:DispersionFunction}.

Thanks to the two expression of the fringe contrast $C_{\Phi}$ and $C_{\Psi}$, including or not the group-delay, we are able to estimate the final fringe contrast on the detector.

This tool has finally two usages:
\begin{itemize}
    \item Selecting the best configuration for the next LDC: we can try many different configurations from a single glass to a combination of mediums located on different spectral bands and compare the final fringe contrasts.
    \item Setting all the media thicknesses during an observation.
\end{itemize}

\subsection{Introducing the transmission loss in the signal-to-noise ratio estimation}

In the previous sections, we demonstrated a formalism for estimating the fringe contrast due to dispersion residues. However, the final signal-to-noise ratio (SNR) of the measurement of the fringe visibilities is impacted both by the dispersion and the transmission of the media. As the media involved in the correction are expected to be transmissive in the considered wavebands, the dispersion is the major contributor to the SNR loss. Yet, to reflect this additional contribution, the transmission loss of the media can be introduced in the formalism. This section demonstrates this refined formalism. The details can be found in appendix~\ref{sec:AppTransmission}.

We define $\Gamma$, the attenuation factor on the SNR:
\begin{equation}
    \Gamma(\Sigma) = T(\Sigma) \cdot C_{\Phi}(\Sigma)^2
    \label{eq:gamma}
\end{equation}
where $T(\Sigma)$ and $C_{\Phi}(\Sigma)$ are respectively the total transmission of the media and the fringe contrast as defined in equation~(\ref{eq:FringeContrast}). The square factor on the fringe visibility comes from the fact that this observable is calculated by averaging the squared modulus of the image Fourier transform. The higher is the $\Gamma$ factor, the higher is the final SNR of the measurement.

The logarithmic attenuation of the transmission $T(\Sigma)$ is defined as:
\begin{equation}\label{eq:Lprime}
  L'=\sum_\Sigma W'_\Sigma\, \boldsymbol{\alpha}_\Sigma \cdot \mathbf{t}^0
\end{equation}
where $\mathbf{t}^0$ is the vector of the average positions of the LDCs, $\boldsymbol{\alpha}=(\alpha_i)_{i\in[1,N]}$ is the vector of the averaged extinction coefficients of the media and $W'_\Sigma$ the weights analog to the $W_\Sigma$ of the dispersion quantity. To respect the contributions of the visibility loss and transmission loss to the SNR, as given by the equation~(\ref{eq:gamma}), these two weights should be linked by the relation
\begin{equation}
    W'_\Sigma = W_\Sigma/2
\end{equation}
such that maximising $\Gamma$ is equivalent to minimising the quantity $L_{tot}$ defined as:
\begin{equation}
    L_{tot} = L + L'
\end{equation}
following equations~(\ref{eq:criterelog}) and (\ref{eq:Lprime}).

\section{Application to CHARA}
\label{sec:solution}
The formalism derived in Sect.~\ref{sec:theory} has been applied to the definition of a chromatism corrector for the CHARA array. In the following subsection, we present the new generation of instruments that will benefit from the new LDC solution and why it is necessary.

\subsection{Presentation of the CHARA instruments}
\label{subsec:CHARAinstruments}

\begin{figure}
    \includegraphics[width=\columnwidth]{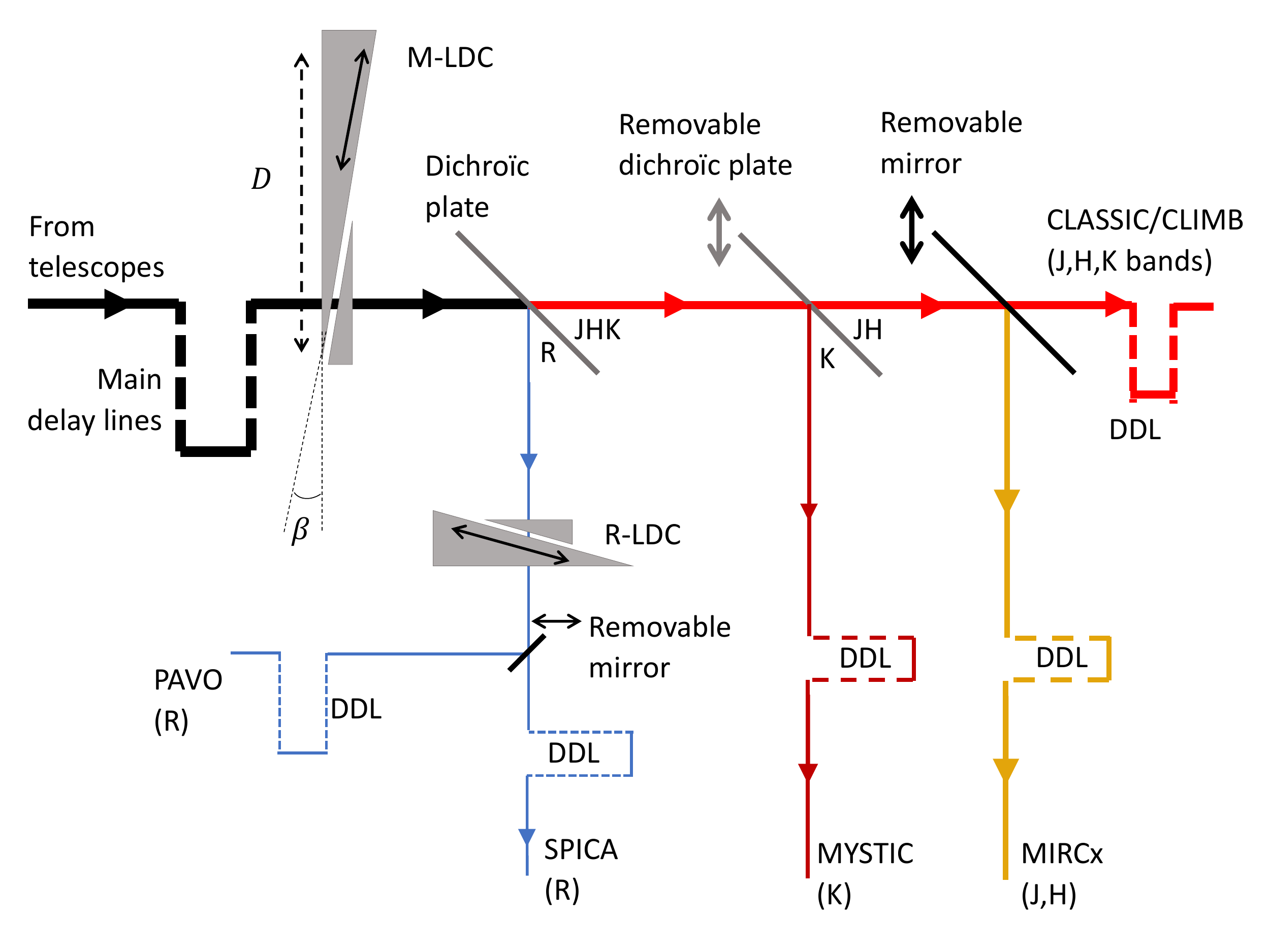}
    \caption{Illustration of the CHARA focal laboratory with its different instruments, their DDLs and the LDCs. This drawing only shows one over the six arms that counts the array.}
    \label{fig:DrawingLDC}
\end{figure}

\begin{figure}
	\includegraphics[width=\columnwidth]{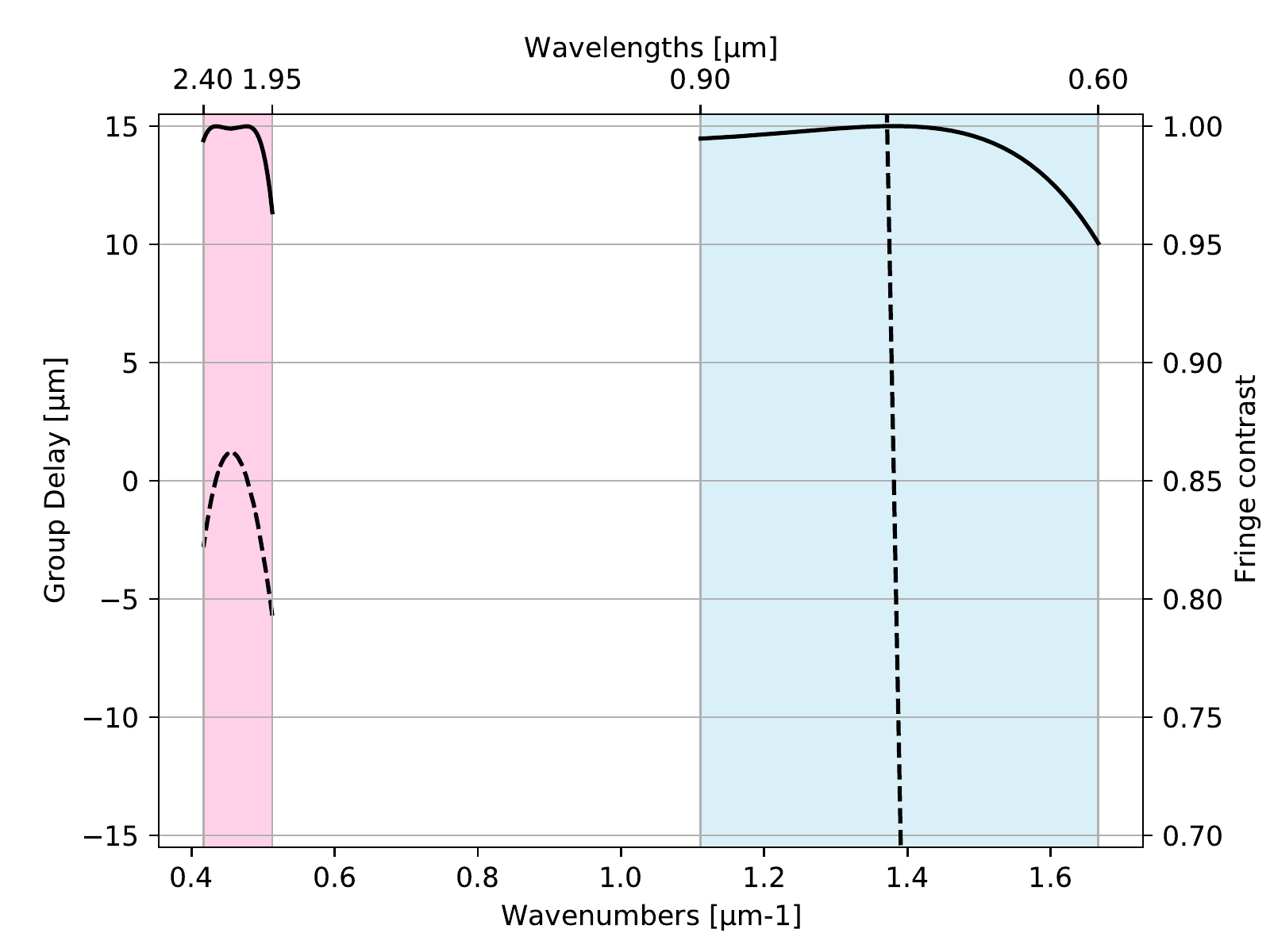}
    \caption{Group-delays (dashed lines) and associated fringe contrasts (solid lines) as seen by the instrument VEGA in the R-band (blue area) and the instrument CLIMB in the K-band (pink area) when the LDC made of SF10 is used along with the differential delay line of CLIMB. As the spectral resolution of VEGA is 6000 and the one of CLIMB is 20, we must favour the correction of dispersion in K band before R-band. That is why the group-delay in the R-band varies very far from the null value but stays close to zero in the K-band.}
    \label{fig:GD_SF10_RK}
\end{figure}

In CHARA, the transportation from the telescope to the lab is done in evacuated pipes but the MDL are in air over their total stroke of 44.5~m. A difference of about 90~m of air appears in the most unfavourable cases, introducing a group-delay of more than 10~mm between the visible fringes and the infrared ones.
For compensating the group delay between the spectral bands R and K, a first LDC was designed by \citet{berger_preliminary_2003}. It consists in two wedged glasses made of SF10 whose total thickness can be tuned by changing their relative positions (see Fig.~\ref{fig:DrawingLDC}). Fig.~\ref{fig:GD_SF10_RK} shows that these LDCs can keep the fringe contrast in the major part of both instrument spectral range above 95\%. But this is done at the price of an important loss of transmission in the infrared bands.

Since 2003, new instruments have been installed on CHARA. The current and coming instruments are gathered in Table~\ref{tab:CHARAfacility}. Most of them include differential delay lines (DDL) in air to equalize the mean group-delay.
In 2018, MIRCx \citep{kraus_mirc-x_2018} has been installed to observe in H-band with the possibility to get data also in J-band. It will be followed by MYSTIC \citep{monnier_mystic_2018} that observes in K-band. A new fringe tracker named SPICA-FT \citep{pannetier_progress_2020} has been plugged into the MIRCx instrument, as an integrated-optics device in H-band and a fast piston controller. It aims at performing fringe-tracking at a frequency of 200~Hz, enabling integrations of~1 to many seconds for all instruments. As for R-band, SPICA-VIS \citep{mourard_spica_2017} is currently in development to replace VEGA~\citep{mourard_performance_2012}.

The fact that these three instruments work at low spectral resolution make the measured fringe contrast more sensitive to the temporal coherence losses due to the longitudinal chromatism induced by the optical delay lines.

In 2019, the CHARA organization considered changing the LDC made of SF10, a glass that suffers from absorption in the infrared, in order to improve the sensitivity in H and K bands.


\begin{table}
    \centering
    \caption{Summary of the configurations of the different instruments in presence on CHARA.}
    \label{tab:CHARAfacility}
    \begin{tabular}{|l||l|l|c|l|} 
     \hline
     Instrument& Year & Waveband & DDL & Resolution\\
     \hline
     CLASSIC/ CLIMB & 2013 & J, H or K &Y& Broadband\\
     MIRCx & 2018 & J, H &Y & 20\\
     MYSTIC & 2018 & K & Y & 20\\
     SPICA-FT & 2021 & H & Y & 20\\
     PAVO & 2008 & 630-900~nm & Y & 30\\
     VEGA & 2007 & 450-850~nm & N & 3000 \\
     SPICA-VIS & 2021 & 600-900~nm & Y & 140\\
     \hline
    \end{tabular}
 \end{table}

\subsection{The LDC configuration for CHARA multi-band observation}
\label{subsec:LDCconfiguration}

The DDLs play the important role of shifting the group-delay of the equipped band to the ZGD position, generally imposed by the fringe tracker in its dedicated spectral band, without introducing any transmission loss. The presence of DDLs on all instruments is thus of great interest for longitudinal dispersion compensation as each one releases one degree of freedom for the LDC to correct the higher order dispersion residues.

To reach the requirements of SPICA in terms of visibility measurements, a fringe contrast better than 95\% must be guaranteed by the LDC in R-band.
The requirements are the same for the H-band where for the fringe-tracker SPICA-FT, to guarantee a precise and fast phase tracking benefiting to all instruments.
We wish also to reach the best dispersion correction possible in K-band to guarantee good performance for MYSTIC. Finally, the instrument MIRCx is now able to record the J and H bands on the same detector~\citep{MIRCX2020} so we must maintain the fringe contrast higher than 95~\% on this whole band to get the best performance from the instrument. Verifying all these conditions, we guarantee fringes observation on R, J, H and K bands simultaneously. Additionally, in the focal laboratory of CHARA, the reserved area for the LDC is limited. By construction, no LDC can be set in the J, H or K bands only and the maximal thickness of the glass is limited to 20~mm for the correction of 90~m of air.

The focal laboratory that hosts the main delay lines and the instruments (Fig.~\ref{fig:DrawingLDC}) is filled with an air under controlled pressure (810~mbar) and temperature (298~K). Its typical relative humidity and its C0\textsubscript{2} content are about 15\% and 450 ppm respectively. The refractive index of the air in this state was modelled from \cite{ciddor_refractive_1996} below 1.5~\micron and \cite{mathar_refractive_2007} above 1.5~\micron. Yet, the refractive index and the group-index of the air, whatever the chosen model, does not deviate by more than $10^{-7}$ from the values given by Ciddor's model. In the worst case, this leads to discrepancies of the group-delay of 10~\micron for 100~meters long delay lines which remains lower than the coherence length of the less resolved instrument. The higher orders of the dispersion models have even less consequences.

Using the Python package ZemaxGlass\footnote{\url{https://github.com/nzhagen/zemaxglass}}, we pickep up the refractive indices of most of the visible and infrared glasses available from the main suppliers (SCHOTT\footnote{\url{https://www.schott.com/english/index.html}}, OHARA\footnote{\url{https://www.oharacorp.com/}}, CDGM\footnote{\url{http://cdgmglass.com/}}), representing 340 glasses in total. Then, using the equation~(\ref{eq:MinimizationSolution}), we optimised the fringe contrast on all configurations involving one or two stage(s) of LDC(s) available with our glass database. However, Table~\ref{tab:1LDC} shows that a single stage of LDC is not sufficient to correct on all bands. Moreover, two stages of LDCs on the common optical path attenuate too much to guarantee a high enough SNR. Finally, the only acceptable solution within our constrain of transmission and available space in CHARA is with one stage of LDC in the main optical path (M-LDC) and another one confined to the R-band optical path (R-LDC), after the first dichroic plate as illustrated on Fig.~\ref{fig:DrawingLDC}. 

Thus, we optimised on all the pairs of glasses present in our database. Table~\ref{tab:GlassComparisons} shows the results for the best of these configurations with their fringe contrast and the associated transmission of the media.

\begin{table}
    \caption{Performance of the two best glasses (SF66 and P-SF68) and the current one (SF10) for simultaneous observations in R, J, H and K bands when the MDL are 90~meters~long. We show also the high dispersion correction performance of ZnS Broad. It is given the excess thickness $x$ (in mm) and the fringe contrasts FC. The transmission is calculated for the maximal excess thickness with the addition of 7~mm owing to the LDC design. The average fringe contrast is calculated on the whole bands.}
    \centering
    \begin{tabular}{|c|c|c|c|c|}
    \hline
    
    M-LDC & SF66 & P-SF68 & SF10 & ZnS Broad \\ \hline
    Excess thickness $x$ &   7   &   10    &   9   &    3      \\ \hline
    
    Average FC  &  0.91    &   0.88     &   0.73   &   0.98 \\ \hline
    
    \begin{tabular}[c]{@{}l@{}}Minimal FC\\ (band)\end{tabular} &   \begin{tabular}[c]{@{}l@{}}0.33\\ (J)\end{tabular}      &    \begin{tabular}[c]{@{}l@{}}0.25\\ (J)\end{tabular}    &  \begin{tabular}[c]{@{}l@{}}0.25\\ (J)\end{tabular}    &   \begin{tabular}[c]{@{}l@{}}0.58\\ (R)\end{tabular} \\ \hline 
    
    \begin{tabular}[c]{@{}l@{}}Transmission\\ 0.75~\micron\end{tabular} &   0.995   &   0.996     &    0.997   &     0.92 \\ \hline
    
    \begin{tabular}[c]{@{}l@{}}Transmission\\ 1.63~\micron\end{tabular} &  0.982  &    0.983    &   0.980   & 0.99  \\ \hline
    
    \begin{tabular}[c]{@{}l@{}}Transmission\\ 2.19~\micron\end{tabular} &  0.87  &    0.87    &   0.875   & 0.99  \\ \hline
    
    \end{tabular}
    \label{tab:1LDC}
\end{table}
 
\begin{table}
\centering
\caption{Performance for simultaneous observations in R, J, H and K bands when the MDL are 90~meters~long and the M-LDC (top line) and the R-LDC (left column) are set to their nominal thickness. The notations and conditions are the same as in Table~\ref{tab:1LDC}. SF66 and S-NPH3 used for the R-LDC give similar performance.}
\label{tab:GlassComparisons}
\begin{tabular}{|l||c|c|c|c|c|}
\hline
R-LDC & M-LDC & SF66 & P-SF68 & SF10 & ZnS Broad \\ \hline \hline
\multirow{11}{*}{SF66} & $x$ (M-LDC)                                                     &   7   &    7    &   15   &     3     \\ \cline{2-6} 
                      & $x$ (R-LDC)                                              &   2   &    2.5    &   3.3   &     1.5     \\ \cline{2-6} 
                      & Average FC                                                    &   0.97   &    0.95    &  0.84    &    0.99      \\ \cline{2-6} 
                      & \begin{tabular}[c]{@{}l@{}}Minimal FC\\ (band)\end{tabular} &   \begin{tabular}[c]{@{}l@{}}0.78\\ (K)\end{tabular}  & \begin{tabular}[c]{@{}l@{}}0.63\\ (K)\end{tabular} &   \begin{tabular}[c]{@{}l@{}}0.63\\ (K)\end{tabular}   &  \begin{tabular}[c]{@{}l@{}}0.95\\ (R)\end{tabular}     \\ \cline{2-6}
                      & \begin{tabular}[c]{@{}l@{}}Transmission\\ 0.75~\micron\end{tabular} &  0.992  &    0.997    &   0.992   &   0.92     \\ \cline{2-6} 
                        &\begin{tabular}[c]{@{}l@{}}Transmission\\ 1.63~\micron\end{tabular} &  0.982  &    0.987    &   0.972   & 0.99  \\ \cline{2-6}
                        &\begin{tabular}[c]{@{}l@{}}Transmission\\ 2.19~\micron\end{tabular} &  0.87  &    0.87    &   0.835   & 0.99  \\ \hline \hline
\multirow{11}{*}{S-NPH3} & $x$ (M-LDC)                                                     &   7   &    10    &   15   &    3      \\ \cline{2-6} 
                      & $x$ (R-LDC)                                              &   1.7   &    2    &   2.7   &    0.5      \\ \cline{2-6} 
                      & Average FC                                                    &  0.97    &    0.95    &   0.84   &    0.99      \\ \cline{2-6} 
                      & Minimal FC                                             &  \begin{tabular}[c]{@{}l@{}}0.78\\ (K)\end{tabular} & \begin{tabular}[c]{@{}l@{}}0.63\\ (K)\end{tabular} & \begin{tabular}[c]{@{}l@{}}0.63\\ (K)\end{tabular}  &  \begin{tabular}[c]{@{}l@{}}0.47\\ (R)\end{tabular}    \\ \cline{2-6} 
                      & \begin{tabular}[c]{@{}l@{}}Transmission\\ 0.75~\micron\end{tabular} &  0.994    &    0.996    &   0.992   &     0.92     \\ \cline{2-6}
                      &\begin{tabular}[c]{@{}l@{}}Transmission\\ 1.63~\micron\end{tabular} &  0.982 &    0.983    &   0.972   & 0.99  \\ \cline{2-6}
                        &\begin{tabular}[c]{@{}l@{}}Transmission\\ 2.19~\micron\end{tabular} &  0.87  &    0.87    &   0.835   & 0.99  \\ \hline
\end{tabular}
\end{table}
 
A complete and easily accessible database of extinction coefficients of all the tested glasses was harder to find than their refractive index. For this reason, we couldn't get an exhaustive ranking of the SNR associated with all configurations, as would permit the $\Gamma$ factor accounting for the SNR attenuation. Instead of that, we got an exhaustive ranking of the dispersion residuals of each configurations after minimization with our code of the respective quantities $L$ (equation~\ref{eq:criterelog}). Then, with the reduced list of the configurations offering the best dispersion correction properties, we used the absorption factor of the glasses to estimate their corresponding SNR attenuation, allowing us to choose the best configuration.

Among the most transmissive optical glasses, SF66 offers the best performance in terms of dispersion compensation, the average fringe contrast being of 97\%, most of the contribution owing to the lower performance in K-band. The two other glasses P-SF68 and S-NPH3 have close performance also. Fortunately, SF66 is also totally transmissive from 0.5 to 1.5~\micron. Its transmission starts getting down in the K-band but the thickness of glass necessary for the correction remains small and enables to keep 87\% of internal transmission up to 2.2~\micron, accounting for the total thickness of the static and mobile prisms.

It can be compared to another solution in Table~\ref{tab:GlassComparisons} that makes use of the infrared material ZnS Broad for the M-LDC. The study has pointed out this same excellent dispersion compensation properties for many infrared glasses but ZnS Broad has the advantage of the stability as it is not hygroscopic. This solution is very interesting for the infrared instruments, since the throughput remains above 99\% along with a close-to-perfect correction of the dispersion on the R, J and H band at the same time. But it costs 8\% of transmission losses in the middle of the R-band when accounting for the total thickness of the static and mobile prisms.

Finally, we chose the solution using SF66 both in M-LDC and R-LDC for two reasons. First, our primary goal is to maximise the sensitivity of SPICA. Second, as discussed later, the low dispersion in H and K bands makes possible synchronized observation without the LDC for saving photons, to the cost of fringe tracking performance. The infrared material ZnS Broad could be the subject of a future completing upgrade where the M-LDC is relocated in the JHK-band (becoming a JHK-LDC) just after the first dichroic of Fig.~\ref{fig:DrawingLDC} and the SF66 replaced with this material.

\subsection{Expected performance}

The chosen configuration can be seen on Fig.~\ref{fig:DrawingLDC} with unrealistic scales. Each LDC is made of two wedged (11.3\degr) windows of SF66 including a static piece of thickness 8~mm. The second piece of the M-LDC is 15,4~mm thick at its maximum to allow the maximal differential thickness of 7,4~mm. The second piece of the R-LDC is 10~mm thick to allow the maximal differential thickness of 2~mm.

In the following subsections, we detail the expected performance of the new LDC solution in the different combinations of instruments on CHARA. The related performance is summarised in Table~\ref{tab:AllPerformances}.

\subsubsection{SPICA-VIS (R-band) with fringe tracking in H-band}

\begin{figure}
\includegraphics[width=\columnwidth]{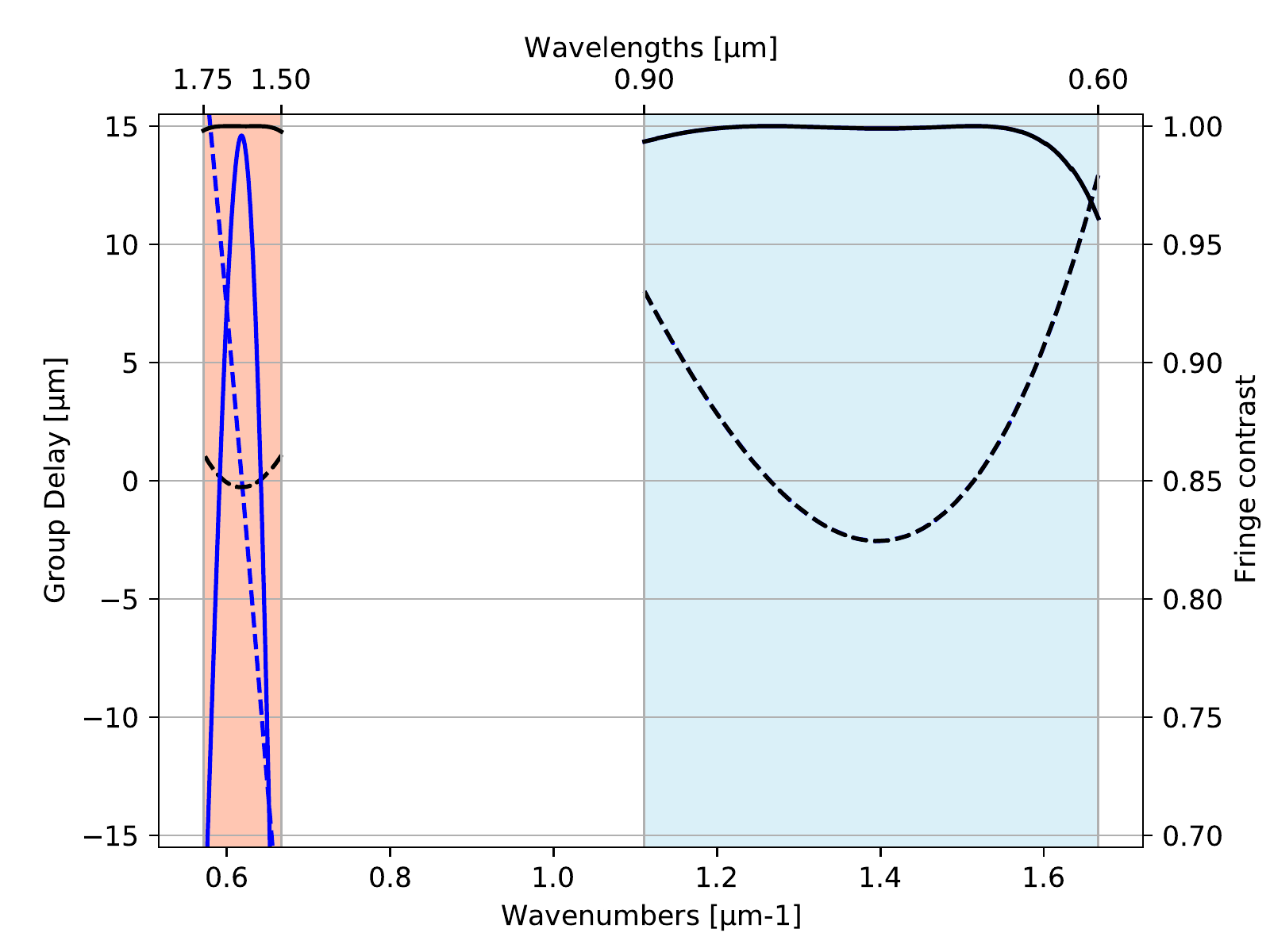}
\caption{Group-delays (dashed lines) and associated fringe contrasts (solid lines) for SPICA-VIS in R-band (blue area) and  MIRCx in H-band (orange area) with the R-LDC and with (black) or without (blue lines) the M-LDC. When there is the M-LDC, MIRCx gets better fringe contrast.}
\label{fig:GD_SF66_RH}
\end{figure}

The Fig.~\ref{fig:GD_SF66_RH} shows the fringe contrast expectation after maximisation in the R and H bands with or without the M-LDC. The R-LDC is always present. The maximal group-delay in the K-band is about 20~\micron (respect. 2~\micron) in absence of M-LDC (respect. in presence of M-LDC) which leads, in case of low spectral resolution R=22, to a fringe contrast below 70\% at the extreme channels whereas it remains over 99\% in presence of the M-LDC. This encourages the use of the two LDCs in this observing mode in order to guarantee a fringe contrast higher than 95\% (respect. 99\%) for SPICA-VIS (respect. SPICA-FT). In addition, we keep an excellent throughput in both bands since it is about 99\% (counting the two LDCs) in R-band and 98\% (only the M-LDC) in H-band, only considering the internal transmission of the glass. Assuming an anti-reflection coating reducing the Fresnel losses to 1\% at each interface, the total transmission losses are around 5\% in R-band and 6\% in H-band.
\NotE{I would like to know the coating performance given by the manufacturer.}

\subsubsection{SPICA-VIS (R-band), MIRCx (J,H bands) and MYSTIC (K-band).}

\begin{figure}
	\includegraphics[width=\columnwidth]{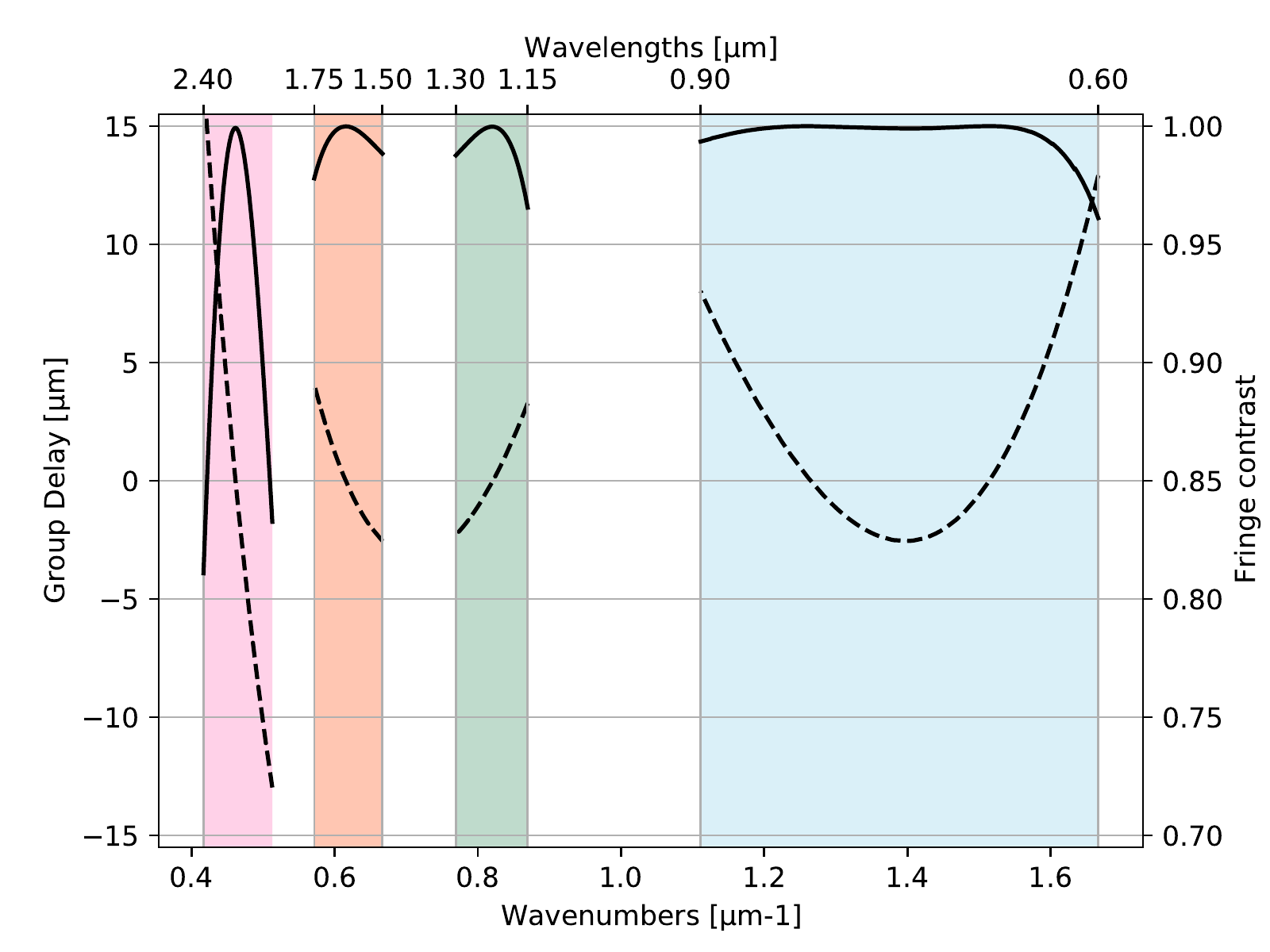}
    \caption{Group-delays (dashed lines) and associated fringe contrasts loss factor (solid lines) in R (blue area), J (green area),  H (orange area) and K (pink area) band with the DDLs of SPICA and MYSTIC and the M-LDC and the R-LDC made of SF66.}
    \label{fig:GD_SF66_RSF66_RJHK}
\end{figure}

\begin{figure}
	\includegraphics[width=\columnwidth]{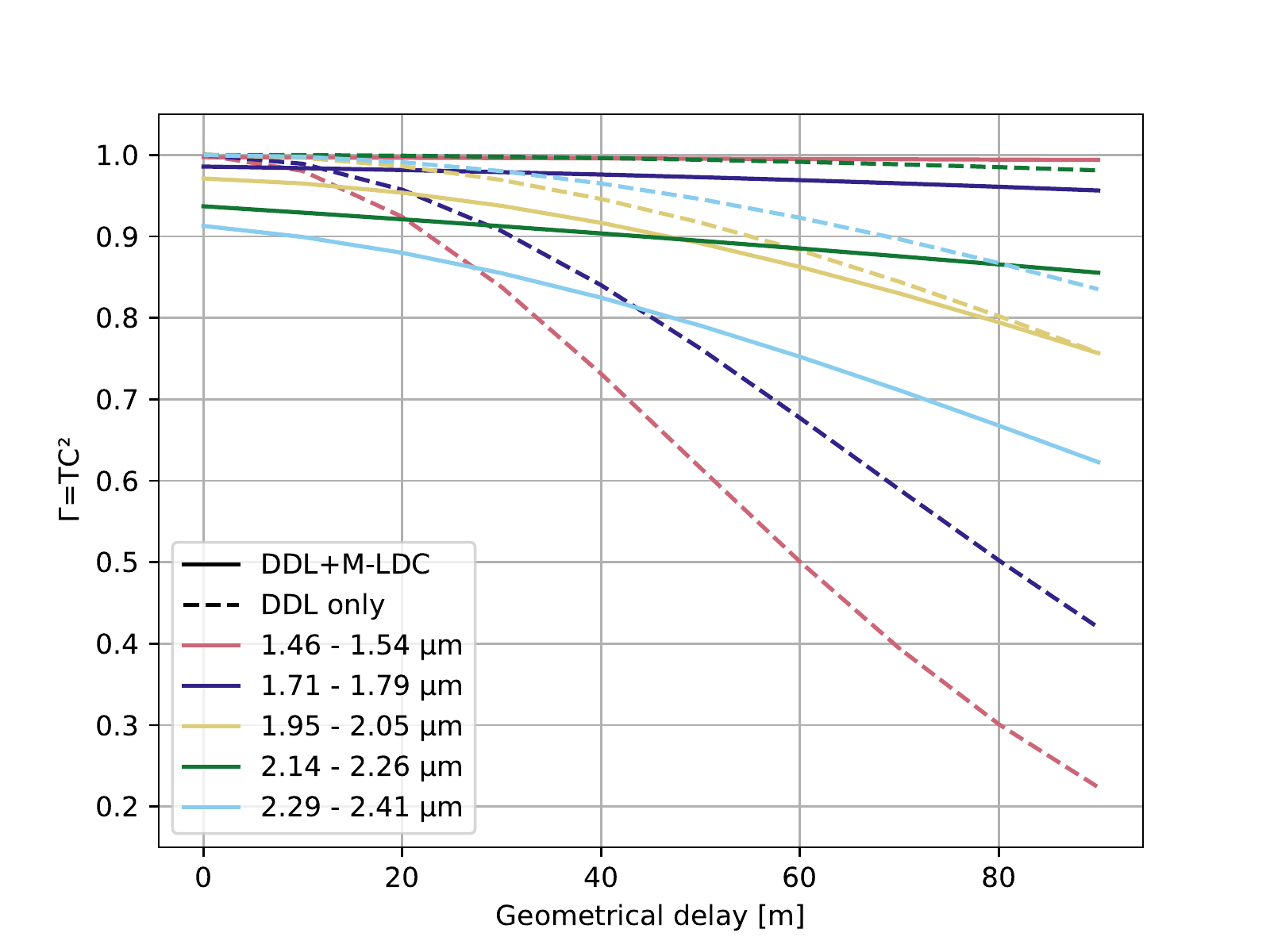}
    \caption{Estimation of the SNR attenuation in the extreme spectral channels in the H and K bands. The middle spectral channel of the K-band is also plotted for additional information. $\Gamma = TC^2$ is plotted as a function of the geometrical delay to be corrected. $\Gamma$ is computed using the transmission of the SF66 at the center of the respective spectral channels and the fringe contrast on these same spectral channels.}
    \label{fig:TVsquare}
\end{figure}

Fig.~\ref{fig:GD_SF66_RSF66_RJHK} shows that the double LDCs solution guarantees a fringe contrast higher than 95\% in R-band and J-band, 97\% in H-band and 80\% in K-band (but higher than 95\% on half the waveband). The high improvement in the three first bands goes with transmission losses in the longest wavelengths, but lower than with the previous LDC. The transmission throughput of the M-LDC in this band (its thickness is equal to 15~mm for 90 meters of MDL, when taking into account its structural minimal thickness of 7.5~mm) is 85\%. Moreover, the Fig.~\ref{fig:GD_SF66_HK} shows that the SF66 doesn't degrade the fringe contrast in this band compared to no LDC at all. Fig.~\ref{fig:TVsquare} shows the evolution of the SNR attenuation with the geometrical delay for the extreme spectral channels in the H and K bands. We focus this figure on these two bands as the glass absorption generates an important attenuation of the SNR in K-band and because the K-band operations are linked to the fringe tracker operation in H-band. 
We see that, without M-LDC, $\Gamma$ falls down in H-band very quickly whereas it stays above 70\% in K-band until 90~m of geometrical delay. With M-LDC, the SNR attenuation remains higher than 95\% in H-band and higher than 63\% in K-band. The high performance in H-band with the M-LDC will permit longer coherent integration for the observing instruments. This will compensate for the transmission loss in K-band and even increase the final sensitivity of the instrument.

\subsubsection{MYSTIC (K-band) and MIRCx (J-band and H-band)}

MYSTIC \citep{monnier_mystic_2018} and MIRCx \citep{kraus_mirc-x_2018} are expected to work alongside during many nights. Both instruments can supply the fringe tracking for the other one, depending on the science goal.

\begin{figure}
\includegraphics[width=\columnwidth]{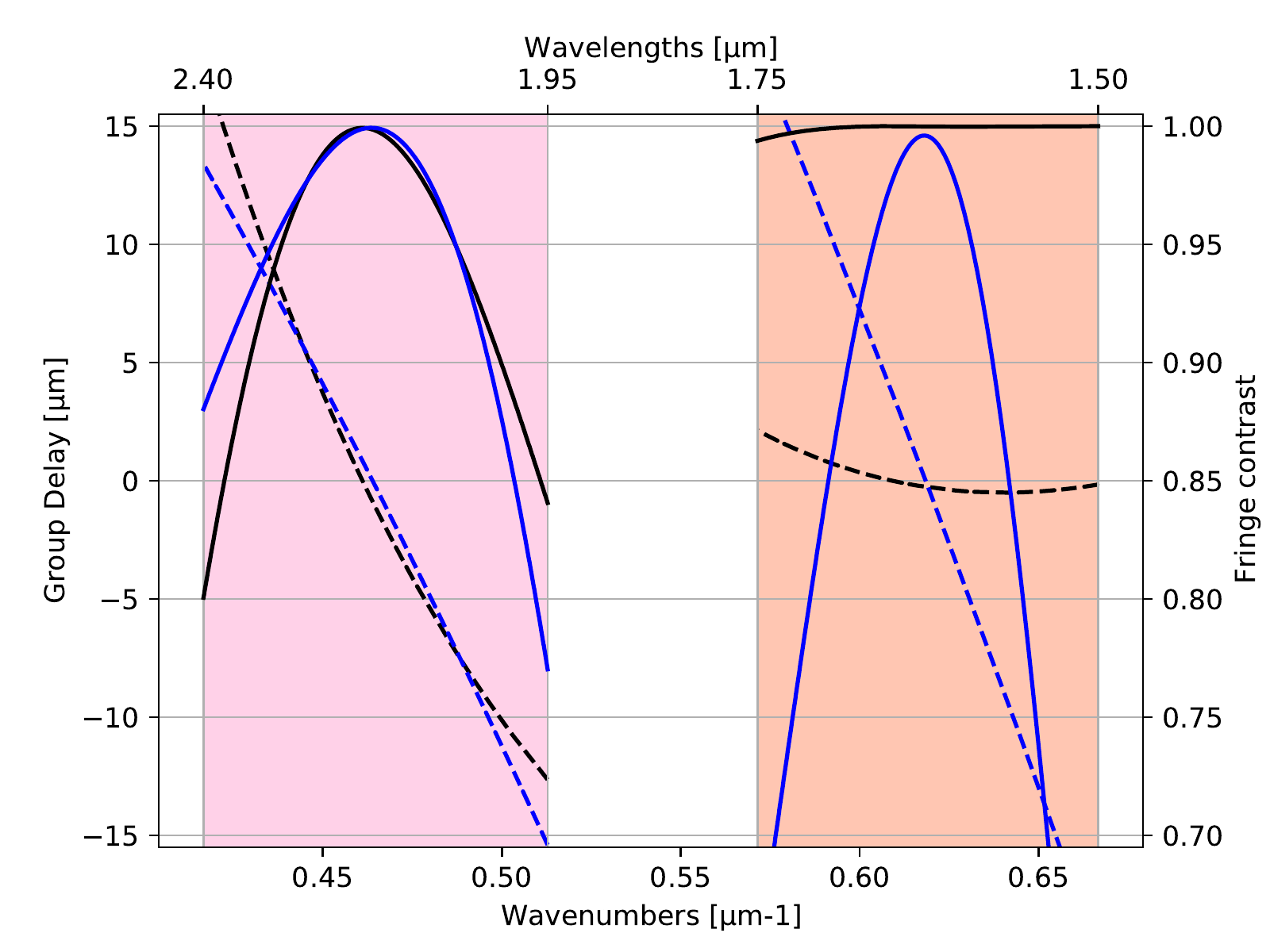}
\caption{Group-delays (dashed lines) and associated fringe contrasts loss factor (solid lines) for MIRCx in H-band (orange area) and  MYSTIC in K-band (pink area) with (black) and without (blue) the M-LDC. MYSTIC's DDL are always used to compensate for the 57~\micron (or 24~\micron with M-LDC) of group-delays between the two instruments. We see that the M-LDC doesn't improve MYSTIC's fringe contrast but improves a lot the fringe contrast of MIRCx.}
\label{fig:GD_SF66_HK}
\end{figure}

In this H-K configuration (see fig.~\ref{fig:GD_SF66_HK}), the M-LDC folds the phase in the H-band such that there is almost no fringe contrast loss on this whole band. The differential delay line nulls the group-delay of MYSTIC.

But the C-RED One detector equipping MIRCx is also sensitive to the J-band and measurements in this band with the two other bands are considered. Due to the configuration of the LDC, this leads to the exact same situation as the R,J,H and K configuration.
The performance in K-band are almost the same than in the H-K configuration with a fringe contrast higher than 80\%.

\subsubsection{PAVO (R-band) with fringe tracking in H-band}

\begin{figure}
\includegraphics[width=\columnwidth]{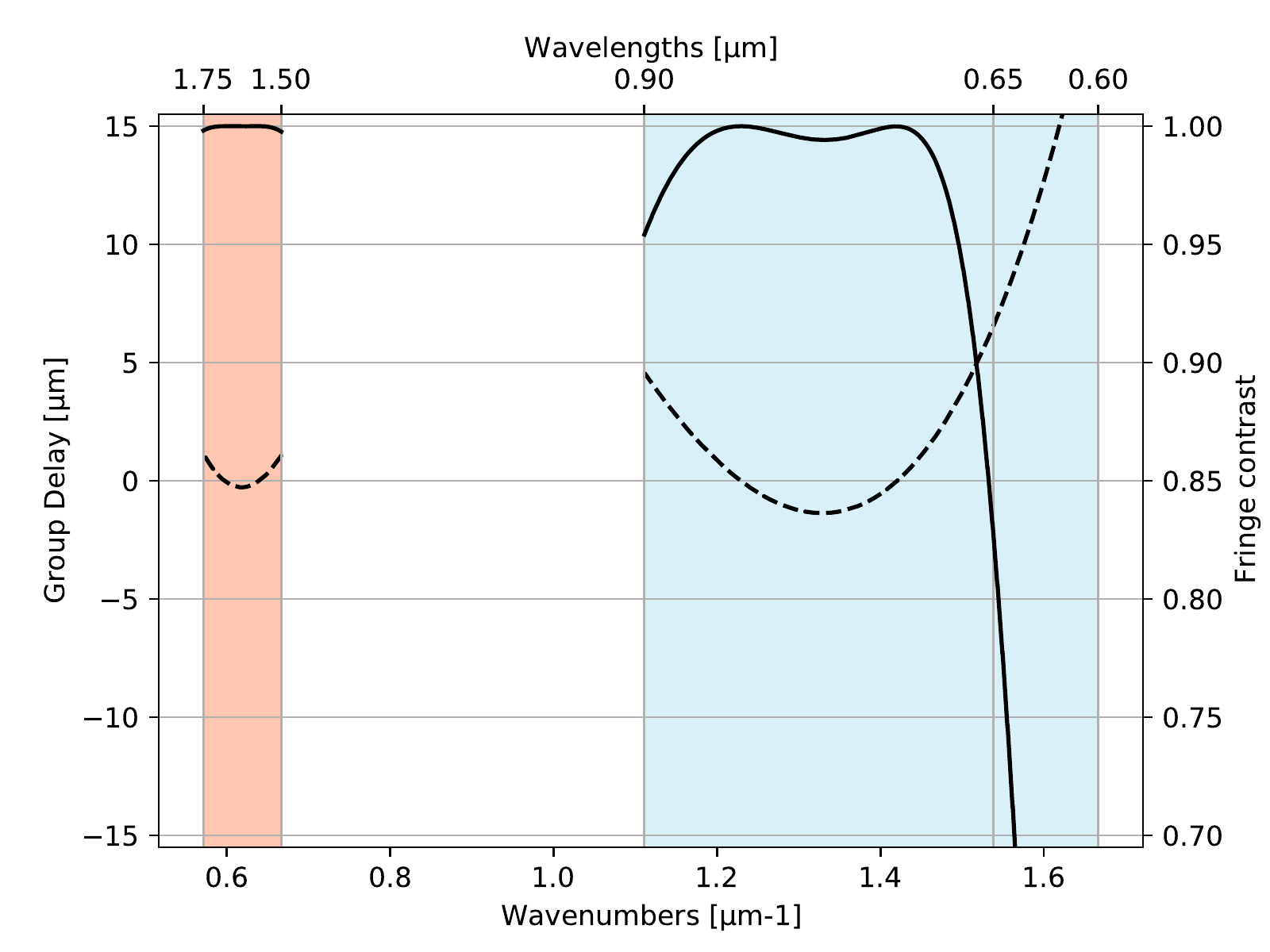}
\caption{Group-delays (dashed lines) and associated fringe contrasts (solid lines) for PAVO in R-band (blue area) and  MIRCx in H-band (orange area) with the new solution optimised for maximisation of the fringe contrast in H-band and between 0.65 and 0.9~\micron.}
\label{fig:GD_SF66_PAVOMIRCx}
\end{figure}

PAVO~\citep{ireland_sensitive_2008} is a temporally modulating spectro-interferometer optimised for high sensitivity in the R-band (spectral resolution 30). Just like SPICA-VIS, it will benefit from the two LDCs and Fig.~\ref{fig:GD_SF66_PAVOMIRCx} shows that the fringe contrast remains high in the whole band while benefiting from a second advantage: the fringe tracking provided by SPICA-FT in H-band.

\begin{table*}
    \centering
    \caption{Summary of the expected performance for the different observing modes for the worst dispersing case corresponding to 90m-long MDLs. When MIRCx is working, it implies also SPICA-FT. R-DDL and K-DDL account respectively for SPICA and MYSTIC's DDL. In the case of PAVO operating with H band, MIRCx's DDL would be used with the excess thickness opposite to the one given in the R-DDL line.}
    \label{tab:AllPerformances}
    \setlength{\tabcolsep}{5pt}    
    \begin{tabular}{|>{\raggedright}p{2.4cm}||c|c|c|c||c|c||c|c|c||c|c||c|c|}
     \hline
     Involved bands or instruments &\multicolumn{4}{c||}{RJHK} & \multicolumn{2}{c||}{HK} & \multicolumn{3}{c||}{JHK}& \multicolumn{2}{c||}{SPICA-VIS,FT} &
     \multicolumn{2}{c||}{PAVO,SPICA-FT}\\
     \hline
     Figure & \multicolumn{4}{c||}{\ref{fig:GD_SF66_RSF66_RJHK}} & \multicolumn{2}{c||}{\ref{fig:GD_SF66_HK}} & \multicolumn{3}{c||}{\ref{fig:GD_SF66_RSF66_RJHK}}& \multicolumn{2}{c||}{\ref{fig:GD_SF66_RH}} & \multicolumn{2}{c||}{\ref{fig:GD_SF66_PAVOMIRCx}}\\
     \hline
     \hline
     \multicolumn{14}{|c|}{Differential thicknesses (\micron)}\\
     \hline
     R-DDL & \multicolumn{4}{c||}{3810} & \multicolumn{2}{c||}{-} & \multicolumn{3}{c||}{-}& \multicolumn{2}{c||}{6664} &
     \multicolumn{2}{c||}{6664}\\
     \hline
     K-DDL & \multicolumn{4}{c||}{-27.4} & \multicolumn{2}{c||}{-23.4} & \multicolumn{3}{c||}{-27.4}& \multicolumn{2}{c||}{-} &
     \multicolumn{2}{c||}{-}\\
     \hline
     M-LDC & \multicolumn{4}{c||}{7320} & \multicolumn{2}{c||}{8319} & \multicolumn{3}{c||}{7322}& \multicolumn{2}{c||}{8831} &
     \multicolumn{2}{c||}{8831}\\ 
     \hline
     R-LDC & \multicolumn{4}{c||}{-1980} & \multicolumn{2}{c||}{-} & \multicolumn{3}{c||}{-}& \multicolumn{2}{c||}{-3488} &
     \multicolumn{2}{c||}{-3488}\\
     \hline
     \hline
     Spectral band & R&J&H&K & H&K & J&H&K & R&H &0.65 - 0.9 & H\\
     \hline
    Spectral resolution & 140&20&20&20 & 20&20 & 20&20&20 & 140&20 & 30&20\\
    \hline
    Absolute max GD [\%~of coherence length] & 12&15&16&37 & 9&40 & 17&16&37 & 12&6 & 56&6\\
    \hline
    Minimal contrast (\%) & 95&97&98&79 & 99&81 & 95&98&79 & 97&99 & 80&99\\
    \hline
    Spectral range where contrast exceeds 95\% & All&All&All&2.1 - 2.3 & All&2.1 - 2.3 & All&All&2.1 - 2.3 & All&All & 0.65 - 0.85&All\\
    \hline
    Mid-band throughput & 0.992 &0.987&0.982&0.87 & 0.982&0.87 & 0.987&0.982&0.87 & 0.992&0.982 & 0.992&0.982\\
    \hline
    Minimal SNR attenuation ($\Gamma$) & 0.90 &0.93&0.94&0.47 & 0.96&0.52 & 0.89&0.94&0.50 & 0.93&0.96 & 0.63&0.96\\
    \hline
    \end{tabular}
\end{table*}

\section{Discussion}
\label{sec:discussion}

The double LDCs solution gives excellent fringe contrast (with high transmission) in the R, J and H bands simultaneously.
Thanks to the dispersion law of SF66, the dispersion in the J and H bands can be very well corrected, enabling to reach fringe contrasts close to 100\% in the whole bands, while not increasing the dispersion residues in the K-band. The fringe contrast in K-band, maximal around its center, remains above 90\% on more than half the spectral channels. The good transmission of SF66 in this band and the small thickness necessary for the correction enables keeping a reasonable throughput higher than 87\% at 2.2~\micron.

\begin{figure}
	\includegraphics[width=\columnwidth]{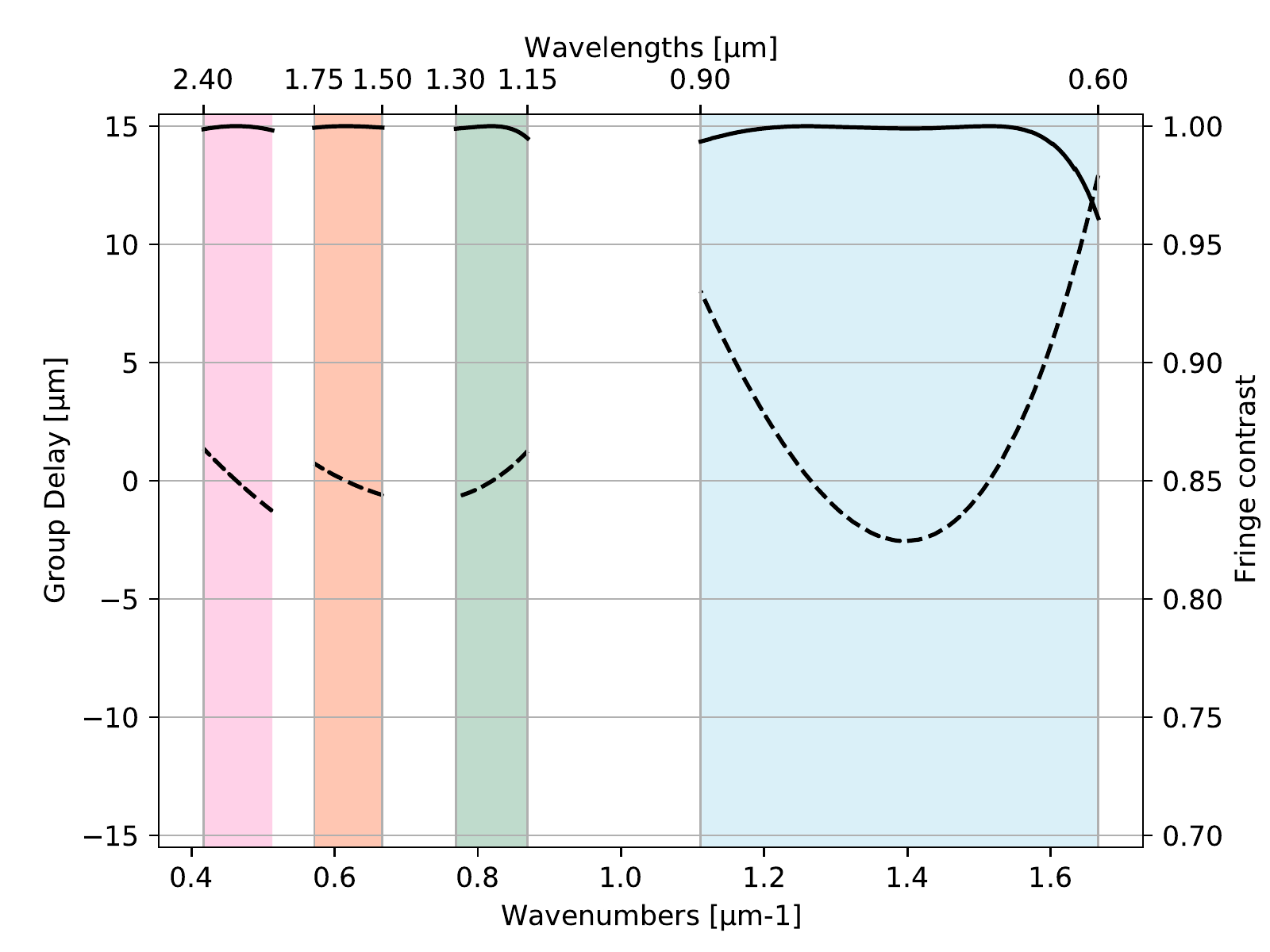}
    \caption{Group-delays (dashed lines) and associated fringe contrasts (solid lines) in R (blue area), J (green area),  H (orange area) and K (pink area) band when the ZnS is introduced in the {J,H,K} bands and SF66 is in the R-band. It implies $x(ZnS)\sim3$~mm and $x(SF66)\sim5$~mm. The same level of performance is observed for infrared glasses such as CsBr, KCl, AgCl, KBr.}
    \label{fig:GD_JHKinfrared_RSF66_RJHK}
\end{figure}

To improve even more the transmission in K-band without impacting the R-band, the only solution we found is to replace the M-LDC by a LDC made of infrared medium like ZnS and located on the JHK optical path. The performance of this solution is close to perfect like we see on Fig.~\ref{fig:GD_JHKinfrared_RSF66_RJHK}. However, two practical constrains prevent us from using it. First, as this glass has a lower transmission in the visible it has to be installed after the infrared/visible dichroic, but the available space on CHARA doesn't allow this at that time. Second, the manufacturers can't guarantee a polishing flatness better than a quarter of a wavelength, which involves wavefront flatness of the same order because of the high refractive index of ZnS. This prevents us from using it after the adaptive optics system, i.e. after the infrared/visible dichroic plate, to keep a high injection factor in the fibered instruments.

\section{Conclusion}

We have proposed a general methodology to address the problem of dispersion compensation in a 2 beams interferometer, resulting in a single matrix equation whose coefficients can be easily computed from instrumental data (index variation of the material involved and bounds of spectral bands). It has been applied to the simultaneous correction of longitudinal dispersion in R, J, H and K bands on CHARA.

We identified SF66 as being the most suitable glass (among standard glass catalogs SCHOTT, OHARA, CDGM and some infrared glasses) for the compensation of the longitudinal chromatism in the visible and near-infrared bands R, J and H while keeping an excellent throughput in K. A low spectral resolution simultaneously on these four bands (R=140 in R-band, R=20 in the three others) is reachable for a differential air thickness up to 90~m with the DDLs and two LDCs: a first one in the common path for maximizing contrast in J and H bands while keeping it high in K-band and a second one in the visible path only for maximizing contrast between 0.6 and 0.9~\micron without degrading the transmission in the K band. With this solution the residual dispersion and the transmission losses of the two LDC stages after correction for 90~meters of MDL are responsible for an attenuation of 0.90, 0.93 and 0.94 on the SNR in respectively R, J and H bands. With the same considerations, the attenuation on the SNR in K-band is 0.47 with our solution and 0.43 in absence of LDC correction (owing exclusively to the dispersion residuals). These values account for the most impacted spectral channels of each spectral band. Since the presence of LDC also increases the SNR for the fringe-tracker in H-band, whose high performance is critical for the instrument MYSTIC in the K band, the use of LDC is clearly benefiting to this instrument.

\section{Acknowledgments}
The CHARA/SPICA instrument is funded by CNRS, Université Côte d'Azur, Observatoire de la Côte d'Azur, and by the Région Sud. The CHARA Array is supported by the National Science Foundation under Grant No. AST-1636624 and AST-1715788.  Institutional support has been provided from the GSU College of Arts and Sciences and the GSU Office of the Vice President for Research and Economic Development. The doctoral fellowship of CP is co-funded by OCA and ONERA.

\section{Data availability}

The data underlying this article will be shared on reasonable request to the corresponding author.



\bibliographystyle{mnras}
\bibliography{article} 




\appendix

\section{Wide-band fringe contrast}
\label{sec:ContrastLoss}

In this section, we detail the calculations that lead to the wide-band fringe contrast given by equation~(\ref{eq:FringeContrast}).

The phase-delay between two arms of the interferometer already defined in the equation~(\ref{eq:PhaseDelay}) is:
\begin{equation}
    \Phi(\sigma)=2\pi\sigma\sum_{i}(n_i(\sigma)x_i)
    \label{eq:AppPhaseDelay}
\end{equation}
We introduce the vector $\mathbf{x}=(x_i)_{i\in[0,N]}$ made up of the geometrical delays and the $N$ media in presence. 
To observe fringes on the detector, we need to introduce a modulation phase $\theta_m$.

For a given wavenumber $\sigma$, the monochromatic interferogram resulting from the phase-delay dispersion and modulation between two coherent beams is:
\begin{equation}
        I(\sigma,\mathbf{x},\theta_m)=\RE{ \bar{I}(\sigma) \left(1+C_{ref}(\sigma) e^{i(\Phi_{\mathbf{x}}(\sigma)+\theta_m(\sigma))}\right)}
\label{eq:AppMonochromaticInterferogram}
\end{equation}
where:
\begin{itemize}
    \item $\Phi_{\mathbf{x}}(\sigma)$ is the phase-delay as given in equation~(\ref{eq:PhaseDelay}) for differential thicknesses $(x_i)_{i\in[0,N]}=\mathbf{x}$ gathered in the vector.
    \item $\theta_m(\sigma)$ is the modulation phase necessary for observing fringes. It can either be introduced spatially (spatial modulation) or dynamically (temporal modulation).
    \item $\bar{I}(\sigma)$ is the incoherent intensity measured on the detector.
    \item $C_{ref}(\sigma)$ is the fringe contrast.
    \item $i$ is the complex number defined as $i^2=-1$.
\end{itemize}


On the detector, each pixel measures the incoherent addition of the monochromatic interferograms at all wavenumbers within the spectral channel. The polychromatic interferogram of a given spectral channel $\Sigma$ is thus only the continuous addition of the monochromatic interferograms given in equation~(\ref{eq:AppMonochromaticInterferogram}).

\begin{equation}
I_\Sigma (\theta_m,\mathbf{x})=\RE{\int_{\Sigma}\bar{I}(\sigma) \left(1+C_{ref}(\sigma) e^{i(\Phi_{\mathbf{x}}(\sigma)+\theta_m(\sigma))}\right)\ds}
\end{equation}
which can be rewritten:
\begin{equation}
    I_\Sigma (\theta_m,\mathbf{x})=\bar{I}_\Sigma+\tilde{I}_\Sigma (\theta_m,\mathbf{x})    
\end{equation}
where
\begin{equation}
    \bar{I}_\Sigma = \RE{\int_{\Sigma}\bar{I}(\sigma)\ds}
\end{equation}
is the polychromatic incoherent intensity on the detector and
\begin{equation}
    \tilde{I}_\Sigma (\theta_m,\mathbf{x})=\RE{\int_{\Sigma}\bar{I}(\sigma)C_{ref}(\sigma) e^{i(\Phi_{\mathbf{x}}(\sigma)+\theta_m(\sigma))}\ds}
\end{equation}
is the coherent term of the interferogram, responsible for the fringe pattern.

For a function f, let's define $\moy{f}_\Sigma $ its normalised weighted mean on a wavenumber set $\Sigma$ with chromatic weights $w(\sigma)$ such that:
\begin{equation}
\moy{f(\sigma)}_\Sigma=\dfrac{\int_\Sigma w(\sigma)f(\sigma)\,d\sigma}{\int_\Sigma w(\sigma)\ds}
\label{eq:AppAverage}
\end{equation}

Now, let's apply it to the function $f(\sigma)=\exp(i\Phi_{\mathbf{x}}(\sigma))$ on the set of wavenumbers $\Sigma$ where the weights $w(\sigma)$ are the product of the detector incoherent illumination $\bar{I}(\sigma)$ and the modulation function $\exp(i\theta_m(\sigma))$:

\begin{equation}
\moy{e^{i\Phi_{\mathbf{x}}(\sigma)}}_{\Sigma}=\dfrac{\int_\Sigma \bar{I}(\sigma)C_{ref}(\sigma) e^{i\theta_m(\sigma)}e^{i\Phi_{\mathbf{x}}(\sigma)} \ds}{\int_\Sigma \bar{I}(\sigma)C_{ref}(\sigma)e^{i\theta_m(\sigma)} \ds}
\label{eq:raw_ratio}
\end{equation}

This is the ratio of the actual interferogram divided by the interferogram in absence of dispersion (i.e. when $\forall \sigma$, $\Phi_{\mathbf{x}}(\sigma)=0$ and thus giving the highest contrast possible). We note this perfect interferogram $\tilde{I}_{0,\Sigma} (\theta_m)$.
\begin{equation}
    \tilde{I}_{0,\Sigma} (\theta_m)=\int_\Sigma \bar{I}(\sigma)C_{ref}(\sigma)e^{i\theta_m(\sigma)} \ds
    \label{eq:PerfectInterferogram}
\end{equation}

Using this new notation, the equation~(\ref{eq:raw_ratio}) gives us:
\begin{equation}
\tilde{I}_\Sigma (\theta_m,\mathbf{x})=\RE{\tilde{I}_{0,\Sigma} (\theta_m) \moy{e^{i\Phi_{\mathbf{x}}(\sigma)}}_\Sigma}
\label{eq:final_ratio}
\end{equation}

Since the goal is to minimize the contrast loss, we expect a small phase residue after minimization. So, for the identification of the best $\mathbf{x}$, we can assume a small dispersion and use the approximation:  
\begin{equation}
\moy{e^{i\Phi_{\mathbf{x}}(\sigma)}}_{\Sigma}\simeq e^{i\moy{\Phi_{\mathbf{x}}(\sigma)}} e^{-\Var[\Sigma]{\Phi_{\mathbf{x}}(\sigma)}/2}.  
\label{eq:MeanApprox}
\end{equation}

A second-order Taylor expansion in Eq.~(\ref{eq:MeanApprox}) would lead to $1-\Var[\Sigma]{\Phi}/2$. This is a classical computation in optics known as the Marechal approximation for the Strehl Ratio \cite{Marechal-47}, but it has been shown later that the formula $\exp(-\Var[\Sigma]{\Phi}/2)$ is much better \cite{Mahajan-83}: the two formulas have the same second-order behavior, but the presence of higher-order terms in the second one considerably enlarges its validity domain. In particular, if $\Phi$ has a Gaussian distribution, then Eq.~(\ref{eq:MeanApprox}) strictly holds. The detailed derivation of Eq.~(\ref{eq:MeanApprox}) can be found in \cite{Ruilier-a-01}.

Finally, this enables us to rewrite the modulated polychromatic interferogram:
\begin{equation}
\tilde{I}_\Sigma (\theta_m,\mathbf{x}) \simeq \RE{\tilde{I}_{0,\Sigma} (\theta_m) e^{i\moy{\Phi_{\mathbf{x}}(\sigma)}_\Sigma}} \cdot C_{\Sigma}(\mathbf{x})
\label{eq:AppFinalInterferogram}
\end{equation}
where:
\begin{itemize}
	\item $\tilde{I}_{0,\Sigma} (\theta_m)$, defined in equation~(\ref{eq:PerfectInterferogram}). It fits in a coherence envelop shaped by the spectrograph's spectral channel shape.
 	\item $\exp(i\moy{\Phi_{\mathbf{x}}(\sigma)}_\Sigma)$ makes appear the phase shift due to the mean phase-delay introduced by the set of media, weighted with the instrument throughput, the spectrum of the source and the monochromatic fringe contrast.
 	\item and an attenuation factor 
    \begin{equation}
        C_{\Sigma}(\mathbf{x})=\exp(-\Var[\Phi]{\Phi_{\mathbf{x}}(\sigma)}/2) \le 1
    \end{equation}
    It is responsible for the attenuation of contrast on the spectral band and includes the losses induced by the mean group-delay and the dispersion.
\end{itemize}
	
Of course, by definition of the modulation $\theta_m$, we want to observe the maximum contrast for $\theta_m=0$.
We can put aside the mean phase difference $\moy{\Phi_{\mathbf{x}}(\sigma)}_\Sigma$ when calculating the contrast $C_{\Phi,\Sigma}$ of the fringes. 
Finally, in the total polychromatic interferogram of a given spectral channel $\Sigma$, the continue part $\bar{I}_\Sigma$ is not impacted by the dispersion effects while the contrast of the modulated part $\RE{\tilde{I}_{0,\Sigma} (\theta_m) \exp(i\moy{\Phi_{\mathbf{x}}(\sigma)}_\Sigma)}$ is reduced by the loss factor $C_{\Sigma}(\mathbf{x})$.

\section{Multi-band fringe contrast maximization equation}
\label{sec:MaximizationEquation}

From the appendix \ref{sec:ContrastLoss}, we know the dependency of the fringe contrast with the dispersion residues.
Let's take a spectro-interferometer observing on several spectral channels $\Sigma$, disjoint or not. We want to maximize the contrast of all the interferograms given by these spectral channels. For doing that, we can minimise $L(\mathbf{x})$ defined as:
\begin{equation}
    L(\mathbf{x})=-\sum_{\Sigma}W_\Sigma \log C_\Sigma(\mathbf{x})
    \label{eq:AppCritereLog}
\end{equation}
where $W_\Sigma$ is the weight arbitrary given to the spectral channel $\Sigma$ to favour an instrument before another.

According to the average defined in equation~(\ref{eq:AppAverage}), the computation of the variance should take into account the source intensity, the fringe contrast and the instrument throughput. For simplicity, we assume all these values unitary on the whole spectral bands.

The equation~(\ref{eq:AppCritereLog}) leads to:
\begin{equation}
    \begin{aligned}
        L(\mathbf{x}) =&2\pi^2\sum_{\Sigma}W_\Sigma\\&\int_{\Sigma}\left[\displaystyle\sum_{i=0}^N \Pi_i(\sigma)(n_i (\sigma)\sigma-\moy{n_i (\sigma)\sigma}_{\Sigma})x_i\right]^2 \ds
    \end{aligned}
\end{equation}

$L$ is quadratic in the space of the differential thicknesses $x_i$.
The minimum of this function of N+1 variables $(x_0,\dots,x_N )$ is found at the position $\mathbf{x}_{opt}$ where the partial derivatives $\dfrac{\partial L}{\partial x_i}$ ($\mathbf{x}_{opt}$) are equal to zero.
These derivatives can be written:
\begin{equation}
    \begin{aligned}
\dfrac{\partial L}{\partial x_i} (\mathbf{x})=&
8\pi^2\displaystyle\sum_{\Sigma}W_\Sigma\\&\int_{\Sigma}(\tilde{n_i}(\sigma)-\moy{\tilde{n_i}}) 
\displaystyle\sum_{j=0}^N x_j (\tilde{n_i}(\sigma)-\moy{\tilde{n_j}}) \ds 
    \end{aligned}
    \label{eq:AppGradientL}
\end{equation}
with $\tilde{n_i}(\sigma)=\Pi_i(\sigma)\sigma n_i(\sigma)$ using the "flag" function
\begin{equation}
    \Pi(\lambda) =
    \begin{cases}
        1 & $if medium $i$ contributes to $\Sigma$,$\\
        0 & $otherwise$
    \end{cases}
\end{equation}

The geometrical delay $x_0$ is a known entrance variable. A first order and trivial correction of the group-delay can be done by setting the ODL to:
\begin{equation}
    x'_1 = \dfrac{-x_0}{n_{g,1}(\sigma_0)}
\end{equation}
such that the group-delay is minimised at an arbitrary wavenumber $\sigma_0$. The refractive index $n_1$ is thus less interesting than the refractive index $n_\epsilon$ defined as:
\begin{equation}
    n_\epsilon(\sigma)=1 - n_1(\sigma)/n_{g,1}(\sigma_0)
\end{equation}
which accounts for the "extra" index of air with respect to vacuum that remains after the correction of the ZGD for the arbitrary wavenumber $\sigma_0$. Note that $\tilde{n}_\epsilon$ follows the same definition introduced for the $\tilde{n}_i$.

This leads us to introduce a new vector $\mathbf{x'} = (\delta x_1, x_2, ..., x_N)$ where $x_1 = x'_1+\delta x_1$ such that $\delta x_1$ corresponds to the (algebraic) excess thickness of ODL necessary for nulling the group-delay.

The equation~(\ref{eq:AppGradientL}) can be rewritten on its vector form that clearly distinguishes the residual dispersion and the correction that needs to be done:
\begin{equation}
    \mathbf{M} \cdot \mathbf{x}'=\mathbf{d}
    \label{eq:AppMatrixDispEquation}
\end{equation}
where:
\begin{itemize}
    \item the vector $\mathbf{d}=(d_i)_{i\in [1,N]}$, made of the covariances between the spectral deviation and the phase dispersion introduced by the additional media, where
    \begin{equation}
        d_i=-x_0 \displaystyle\sum_{\Sigma}W_\Sigma\int_{\Sigma} (\tilde{n}_{\epsilon}(\sigma)-\moy{\tilde{n}_{\epsilon}})(\tilde{n_i}(\sigma)-\moy{\tilde{n_i}})\ds
        \label{eq:DispersionVector}
    \end{equation}
    \item the matrix $\mathbf{M}=(m_{ij})_{(i,j)\in[1,N]^2}$, made of the covariances between all additional media's phase dispersion, where 
    \begin{equation}
        m_{ij}=\displaystyle\sum_{\Sigma}W_\Sigma\int_{\Sigma}(\tilde{n_i}(\sigma)-\moy{\tilde{n_i}})(\tilde{n_j}(\sigma)-\moy{\tilde{n_j}})\ds
    \end{equation}
\end{itemize}

The equation \ref{eq:AppMatrixDispEquation} is sensitive to numerical noise due to the fact that the matrix $\mathbf{M}$ is ill-conditioned. For the best corrector, its conditioning is typically between $10^{-6}$ and $10^{-8}$. It gets higher as the spectral band to correct gets thinner. Indeed, the majority of the correction is done by putting the average group-delay to zero. However, albeit ill-conditioned, the inverse $\mathbf{M}^{-1}$ of the matrix $\mathbf{M}$ still exists and this equation admits only one solution
\begin{equation}
    \mathbf{x}'_{opt} = \mathbf{M}^{-1} \cdot \mathbf{d}
    \label{eq:UniqSolution}
\end{equation}
that corresponds to the control equation for the $N$ media that form the dispersion control in addition to the first order correction $x'_1$.

\section{Accounting for the transmission}
\label{sec:AppTransmission}
For a wide band LDC, glasses with good transmission all across the band are difficult to find. This appendix shows how to add transmission losses in the dispersion criterion from appendix~\ref{sec:MaximizationEquation} to help identifying the best glasses for the LDC.

For a 2-beam interferometer, the vector of LDC thicknesses on arms~1 and~2, $\mathbf{t}^1$ and $\mathbf{t}^2$ respectively, can always be written as
\begin{equation}\label{eq-LDCcommand}
    \mathbf{t}^a=\mathbf{t}^0+\epsilon^a\,\mathbf{x}/2
\end{equation}
where $a\in\{1,2\}$ is the arm index, $\epsilon^a=-(-1)^a$ and $\mathbf{t}^0$ is the common vector of average positions, which does not impact on the visibility since only the differential delays $\mathbf{x}$ between the arms matter. This offset $\mathbf{t}^0$ results from the LDC geometry (cf fig.~\ref{fig:DrawingLDC}) and the necessity to always have positive thicknesses despite the fact that $\mathbf{x}$ can have both positive and negative values for maximum sky coverage. 



The global transmission $\tau_a$ on each arm $a$ is thus
\begin{equation}
  \tau^a_\Sigma\simeq\exp(-\boldsymbol{\alpha}_\Sigma\cdot\mathbf{t}^a),
\end{equation}
where $\boldsymbol{\alpha}_\Sigma$ is the vector of absorption coefficients at the central wavenumber in each (small) channel. 

Maximizing the fringe attenuation factor, proportional to $\sqrt{\tau^1_\Sigma\,\tau^2_\Sigma}$ in each channel, boils down to minimizing a new criterion $L'$, in logarithmic scale as the visibility attenuation in Eq.~(\ref{eq:criterelog}), which writes:
\begin{equation}\label{eq:AppLprime}
  L'=\sum_\Sigma W'_\Sigma\, \boldsymbol{\alpha}_\Sigma \cdot \mathbf{t}^0
\end{equation}
where $W'_\Sigma$ are weights, analogous to those introduced for dispersion.

For a 2-beam interferometer, $L'$ is only affected by $\mathbf{t}^0$ and not by $\mathbf{x}$, since when reducing thickness on one arm, the same length is added on the other arm with the symmetric command of Eq.~(\ref{eq-LDCcommand}). To minimize the transmission loss, $\mathbf{t}^0$ has to be minimized at each command, such that 
\begin{equation}\label{eq:cmde-optimale}
    \min_a(t_i^a)=t^0_{i,\min}
\end{equation}
where $a$ runs over all the sub-aperture indexes and the minimum value $t^0_{i,\min}$ for each $t_i^0$ results from the diameter $D_i$ and angle $\beta_i$ of the $i$\textsuperscript{th} LDC prism (Fig.~\ref{fig:DrawingLDC}):
\begin{equation}
    t_{i,\min}^0=D_i\,\tan\beta_i.
\end{equation}

The angle $\beta_i$ is a free LDC parameter, only constrained by the maximum correction to apply $|x_{i,\max}|$ and the stroke $B_i$ of the $i$\textsuperscript{th} translation stage. Its minimum value is 
\begin{equation}\label{eq:betaOpt}
    \beta_{i,\min}=\arctan\left(\frac{|x_{i,\max}|}{B_i}\right).
\end{equation}




With more than two beams, control equations similar to Eqs.~(\ref{eq-LDCcommand}) and (\ref{eq:cmde-optimale}) can be derived.
But for simplicity, the criterion $L'$ can be computed only for the baseline leading to the largest $|\mathbf{x}|$, $|\mathbf{x}_{\max}|$, which dominates performance.

The effect of intensity losses, assuming optimal design and equal diameters (D) and stroke (B) for all LDC glasses, is thus bounded by
\begin{equation}
    L'(\mathbf{x}_{\max})=\left(\frac12+\frac{D}{B}\right)\,\sum_\Sigma W'_\Sigma\, \boldsymbol{\alpha}_\Sigma\cdot|\mathbf{x}_{\max}|.
\end{equation}


This term can be used to compute a total criterion $L+L'$ taking the LDC transmission into account. 



\bsp	
\label{lastpage}
\end{document}